\documentclass[usenatbib]{mn2e}
\usepackage[usenames,dvipsnames,svgnames,table]{xcolor}
\usepackage{hyperref}
\definecolor{darkblue}{rgb}{0.0,0.0,0.3}
\hypersetup{colorlinks,breaklinks,
            linkcolor=darkblue,urlcolor=darkblue,
            anchorcolor=darkblue,citecolor=darkblue}
\usepackage{amsmath,amssymb}

\usepackage{txfonts}
\DeclareSymbolFont{cmletters}{OML}{cmm}{m}{it}
\DeclareMathSymbol{v}{\mathalpha}{cmletters}{"76}

\usepackage{graphicx}

\defcitealias{mtb12b}{MTB12a}
\newcommand{\acknowledgements}{\section*{Acknowledgements}}

\newcommand{\be}{\begin{equation}}
\newcommand{\ee}{\end{equation}}

\newcommand{\Rlc}{R_{\rm LC}}
\newcommand\simless\lesssim   
\newcommand\simmore\gtrsim   

\newcommand\avg[1]{{\langle #1 \rangle}}
\newcommand\p{\partial}
\usepackage{macros}

\begin{document}
\label{firstpage}

\title[Three-dimensional Pulsar Wind]{Three-dimensional Analytical Description of Magnetised Winds from Oblique Pulsars}
\author[]{
Alexander Tchekhovskoy$^{1,2}$\thanks{Einstein Fellow}\thanks{Email:
  atchekho@berkeley.edu, philippo@astro.princeton.edu, anatoly@astro.princeton.edu}, Alexander Philippov$^{3}$ and Anatoly Spitkovsky$^{3}$\\
$^1$Department of Astronomy, University of California Berkeley, Berkeley,
  CA 94720-3411\\
$^2$Lawrence Berkeley National Laboratory, 1 Cyclotron Rd,
  Berkeley, CA 94720, USA\\
$^3$Department of Astrophysical Sciences, Peyton Hall, Princeton University, Princeton, NJ 08544, USA}

\date{Accepted . Received ; in original form }
\pagerange{\pageref{firstpage}--\pageref{lastpage}} \pubyear{2015}
\maketitle

\begin{abstract}
Rotating neutron stars, or pulsars, are plausibly the source of power
behind many astrophysical systems, such as gamma-ray bursts,
supernovae, pulsar wind nebulae and supernova remnants.  In the past
several years, 3D numerical simulations made it possible to compute
pulsar spindown luminosity from first principles and revealed that
oblique pulsar winds are more powerful than aligned ones. However,
what causes this enhanced power output of oblique pulsars is not
understood. In this work, using time-dependent 3D
magnetohydrodynamic (MHD) and force-free simulations, we show that,
contrary to the standard paradigm, the open magnetic flux, which
carries the energy away from the pulsar, is laterally
non-uniform. We argue that this non-uniformity is the primary reason
for the increased luminosity of oblique pulsars. To demonstrate
this, we construct simple analytic descriptions of aligned and
orthogonal pulsar winds and combine them to obtain an accurate 3D
description of the pulsar wind for any obliquity. Our approach
describes both the warped magnetospheric current sheet and the
smooth variation of pulsar wind properties outside of it. We find
that generically the magnetospheric current sheet separates plasmas
that move at mildly relativistic velocities relative to each other.
This suggests that the magnetospheric reconnection is a type of driven,
rather than free, reconnection. The jump in magnetic field
components across the current sheet decreases with increasing
obliquity, which could be a mechanism that reduces dissipation in
near-orthogonal pulsars.  Our analytical description of the pulsar
wind can be used for constructing models of pulsar gamma-ray
emission, pulsar wind nebulae, and magnetar-powered core-collapse
gamma-ray bursts and supernovae.
\end{abstract}


\section{Introduction}

A rotating magnet in vacuum is the simplest description of a rotating
neutron star, or pulsar. The magnet loses its energy to magnetodipole
radiation \citep{deutsch55}. However, real pulsar magnetospheres are likely filled with
plasma \citep{gol69}, which supports charges and currents that modify
the pulsar spindown. For a long time it has been realized
that plasma-filled magnetospheres open up to infinity due to relativistic rotation
\citep*{1973ApJ...186..625I,1974ApJ...187..585M}.  The open magnetic
flux carries away the spindown luminosity of the star in the form of
a pulsar wind.

Whereas the origin of spindown luminosity of vacuum rotators has been
well-understood due to the availability of the analytic model, the
spindown of plasma-filled magnetospheres can only be studied
numerically, and progress became possible only recently.  The
primary difficulty in computing pulsar magnetosphere structure is in
connecting the conditions on the star to the properties of the
emerging pulsar wind.  Of particular importance is the determination
of the twist of magnetic field lines at the stellar surface, or the
magnitude of the electric current flowing through the magnetosphere.
Early pulsar magnetosphere solutions prescribed the magnitude of the
twist, 
and, as a result, the solutions extended smoothly only to the
\emph{light cylinder} radius \citep*{BGI,1994MNRAS.271..621M,bes97},
the distance from the rotational axis at which a field line rigidly
rotating at the frequency of the central object would move at the
speed of light, 
\begin{equation}
  \label{eq:Rlcdef}
  \Rlc = \frac{c}{\Omega},
\end{equation}
where $\Omega=2\pi/P$ is pulsar angular frequency and $P$ is the period.

In real magnetospheres, the magnitude of the magnetospheric current is
set self-consistently by the large-scale properties of the
magnetosphere.  The rotation of the star twists the magnetic field
lines, which sends out a torsional Alfv{\'e}n wave that carries the
twist outward.  On the way out, the Alfv{\'e}n wave undergoes partial
reflections that reach back to the star and modify the surface value
of the magnetic field twist, which determines the strength of the
magnetospheric current. It is accounting for this
communication by Alfv{\'e}n waves and determining the correct
value of the magnetospheric current that was the
challenge.
\citet*{ckf99} found an iterative way to determine the strength of the
magnetospheric current self-consistently. They obtained the
first solution for an \emph{aligned} force-free pulsar magnetosphere
that extended out to infinity. Their results were subsequently
verified by other groups within force-free and magnetohydrodynamic
(MHD) approximations (e.g.,
\citealt*{gruzinov_pulsar_2005,tim06,mck06pulff,komi_ns_06,par12phaedra,2014PhRvD..89h4045R})
as well as using a particle-in-cell (PIC) approach
\citep*{2014ApJ...785L..33P,ChenPIC,2014arXiv1410.3757C,2014arXiv1412.2819B}. 

\citet{spit06} carried out the first 3D, oblique pulsar magnetosphere
simulations. Using the force-free approximation, he found that pulsar
spindown luminosity increases with increasing obliquity angle,
$\alpha$, the angle between the rotational and magnetic
axes. More recently, these results were confirmed using time-dependent
3D force-free \citep*{kc09,petri12a,kalap12}, MHD \citep*{SashaMHD}, and
PIC \citep*{2014arXiv1412.0673P} studies.  

Despite these advances, what sets the spindown luminosity of plasma-filled pulsars
remains elusive.  
Understanding this and constructing a detailed, 3D model of a pulsar
wind are the goals of this work.  We start with a discussion of pulsar
spindown in the split-monopole approximation and confront it with 
the numerical simulations in \S\ref{sec:dependence-spin-down},
discuss the effects of pulsar wind non-uniformity on pulsar spindown
rate and develop a toy model of a non-uniform pulsar wind in
\S\ref{sec:non-unif-puls}. We construct an analytic description of
aligned and orthogonal plasma-filled pulsar winds in
\S\ref{sec:appr-solut-align}. We then try and get an inspiration on how to construct
solutions for intermediate inclination angles in
\S\ref{sec:puls-wind-struct} and construct our analytical description
of the pulsar wind for an arbitrary inclination angle in
\S\ref{sec:semi-analyt-solut}. We discuss our results and their
observational implications in \S\ref{sec:disc-concl}. We will use spherical
polar coordinates, $(r,\theta,\varphi)$, cylindrical
coordinates, $(R,z,\varphi)$, and Cartesian coordinates, $(x,y,z)$.

\section{Dependence of spin-down power on inclination and the
  structure of pulsar wind}
\label{sec:dependence-spin-down}

\begin{figure}
  \centering
  \includegraphics[width=0.9\columnwidth]{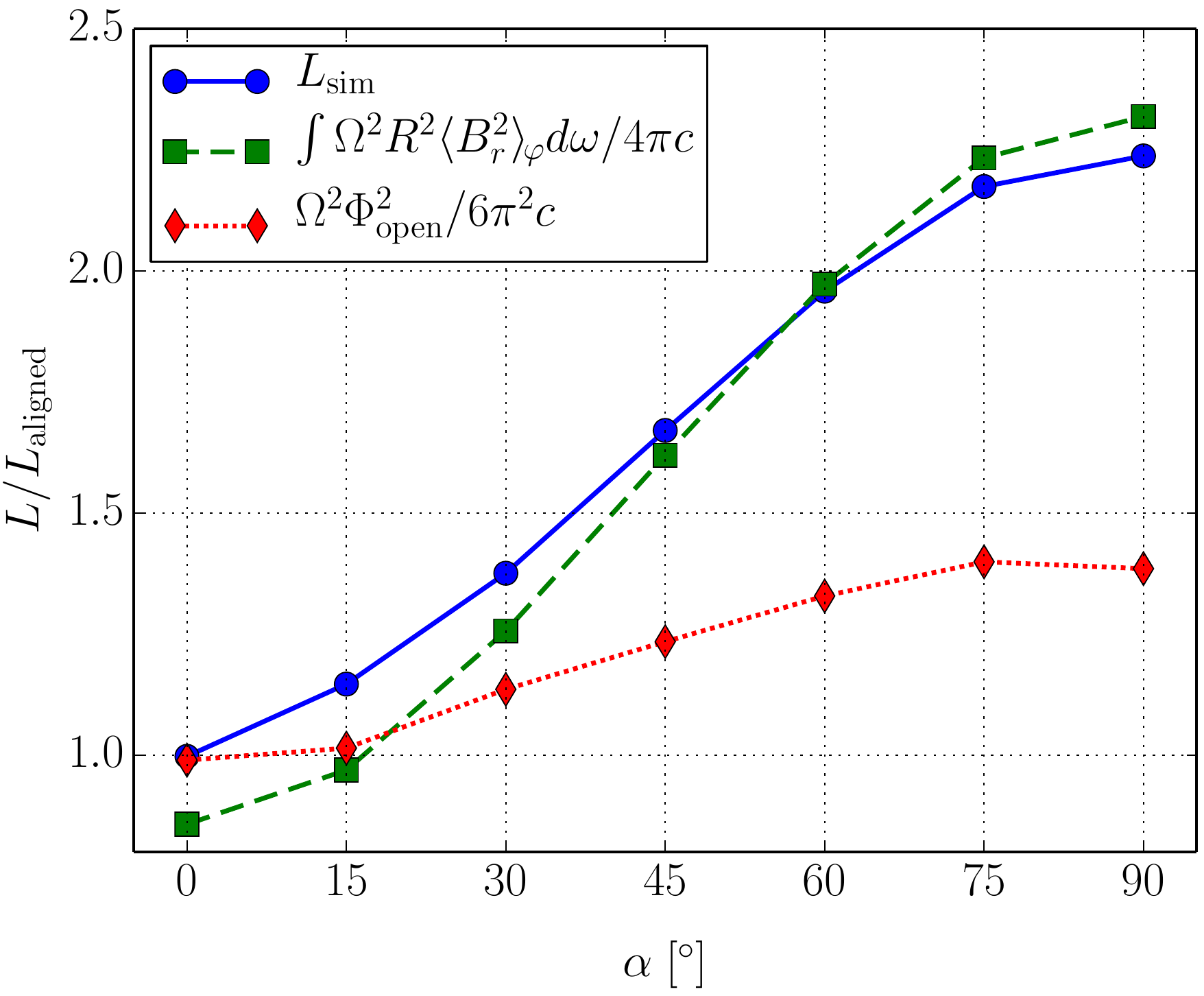}
  \caption{Blue solid line shows the dependence of spindown luminosity
    on the inclination angle, $\alpha$, in a series of MHD simulations
    \citep{SashaMHD}. Red dotted line shows the corresponding spindown
    luminosity for a split-monopole wind (same open magnetic flux as
    in the simulation). It accounts for about $40\%$ of the increase
    but falls short of explaining the full magnitude of the increase
    in luminosity with increasing $\alpha$. Green dashed line shows
    the spindown for a non-uniform distribution of open magnetic flux
    (same non-uniformity as in the simulation). It matches the
    simulated luminosity well. The non-uniformity of the
    open magnetic flux plays a crucial role in setting the spindown
    luminosity of oblique pulsars.}
  \label{fig:Lvsalpha}
\end{figure}

\begin{figure}
 \centering
 \includegraphics[width=1\columnwidth]{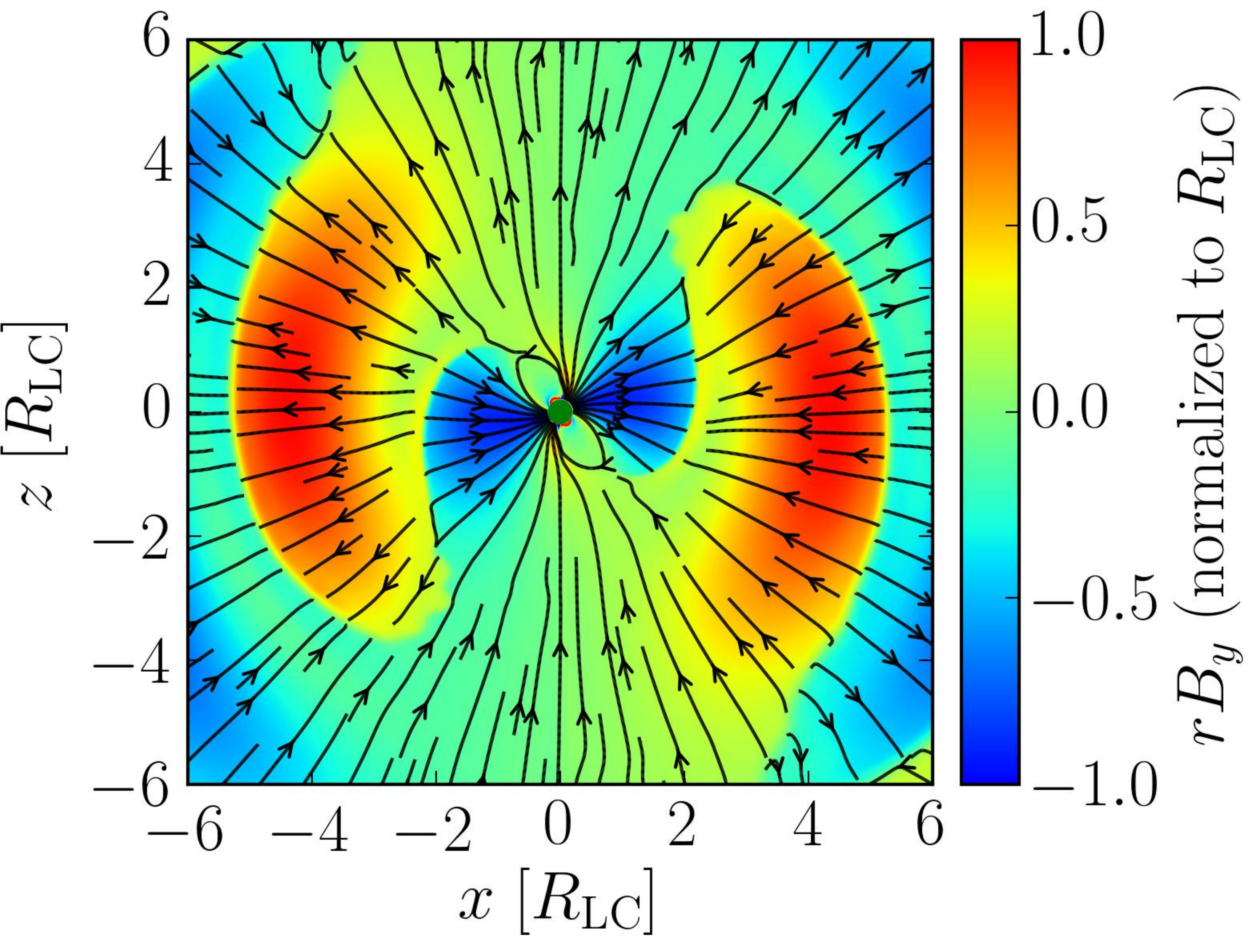}
 \caption{Magnetic field structure of pulsar wind in the
   $\vec\mu{-}\vec\Omega$ plane of an oblique,
   $\alpha=60^\circ$, MHD simulation. Solid black lines show magnetic
   field lines and color shows $rB_y$, into-the-plane magnetic field
   component (which we rescaled by the $r$ prefactor to remove the
   trivial radial trend). That the magnetic field lines are
   predominantly straight suggests that the magnetospheric outflow 
   is nearly radial. However, as indicated by non-trivial
   variations of $rB_y$, whose absolute magnitude would be a constant along
   $\theta=$~constant in a split-monopole solution, this does not mean
   that the solution is well-described by a split-monopole (see also
   Fig.~\ref{fig:pwindBr}). }
 \label{fig:pwind}
\end{figure}

Our goal is to understand the physics of spindown of plasma-filled
pulsars. The solid blue line in Fig.~\ref{fig:Lvsalpha} shows the
dependence of pulsar spindown on
inclination (see \citealt{spit06,petri12a,SashaMHD,2014arXiv1412.0673P}). 
Clearly, oblique pulsars have higher spindown rates, but
the reason for this increase is unclear. In the widely used and highly
successful standard model of the asymptotic pulsar wind, the wind structure is radial
and is that of a split-monopole outflow \citep{Bogovalov}. The wind indeed
appears to be radial far from the pulsar, as is seen in Fig.~\ref{fig:pwind}.
However, this brings up a problem: the spindown of split-monopole
solution is \emph{independent} of inclination! In it,
$\left|r^2B^r\right| = {\rm constant}$ and is time-independent. In fact, as
we will derive below (see eq.~\ref{eq:Pdip}), the spindown luminosity
is set by the rotational frequency of the star,
$\Omega_* = 2\pi/P$, and the amount of open magnetic flux in the
split-monopole solution,
\begin{equation}
\Phi_{\rm open} = (1/2)\iint |B_r|d\omega,
\label{eq:phiopen}
\end{equation}
where the integral is over both hemispheres and the factor of $(1/2)$
converts it to a single hemisphere. For a split-monopole magnetic
field, eq.~\eqref{eq:phiopen} reduces to $\Phi_{\rm open} = 2\pi r^2|B_r|$,
and the spindown luminosity is
\begin{equation}
  \label{eq:Pmono}
  L = \frac{\Omega^2 \Phi_{\rm open}^2}{6\pi^2 c}, \quad ({\rm
    uniform}\ |r^2B_r|)
\end{equation}
where $c$ is the speed of light.

\begin{figure}
  \centering
  \includegraphics[width=1\columnwidth]{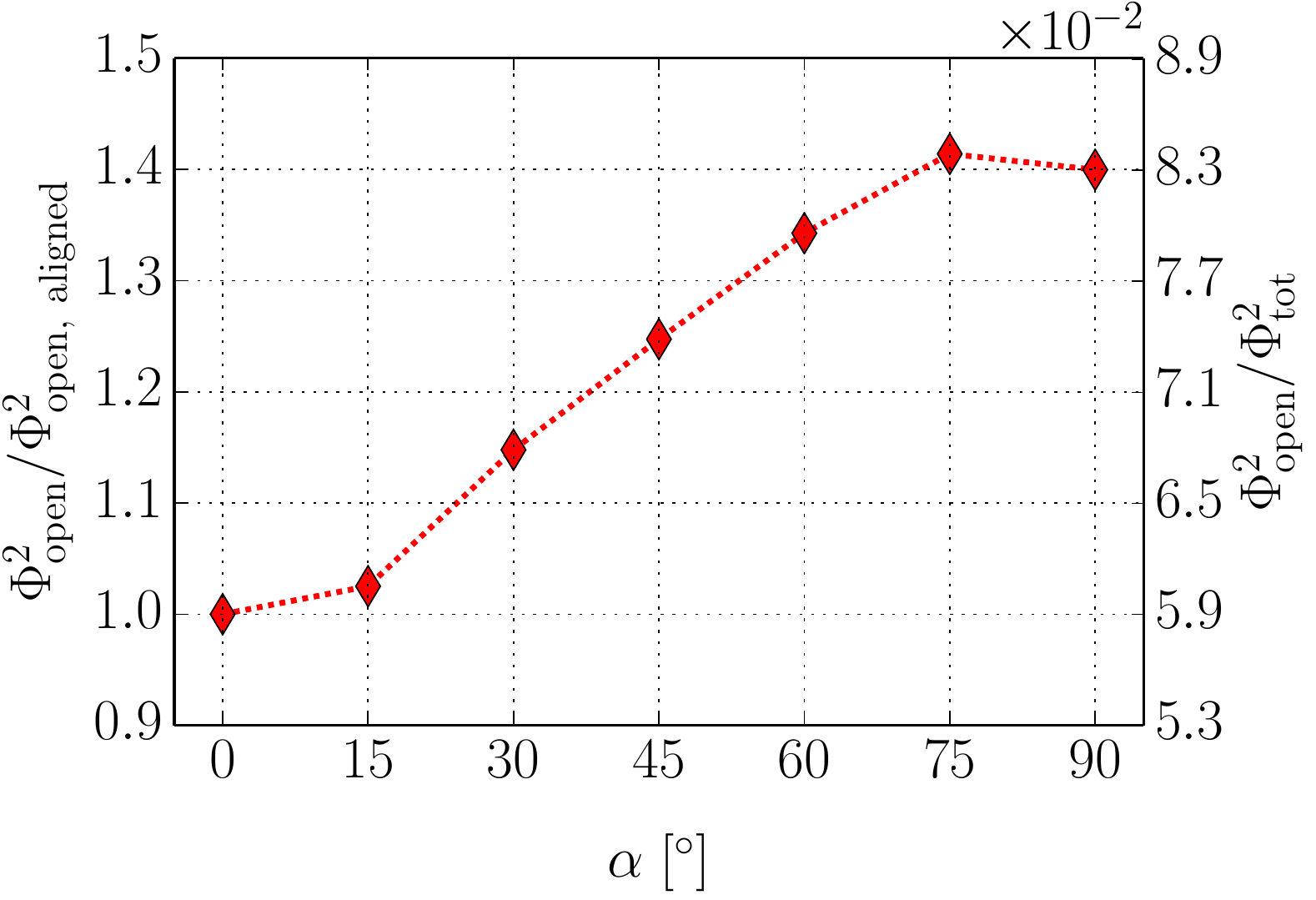}
  \caption{The square of the open magnetic flux $\Phi_{\rm
      open}$ measured according to eq.~\eqref{eq:phiopen} at a sphere of radius
    $r = 2\Rlc$, plotted versus the inclination angle, 
    $\alpha$, in a series of MHD simulations. 
    The open magnetic flux increases with increasing inclination
    angle. This increase accounts for $\approx40\%$ of  the
    enhancement in spindown luminosity of oblique rotators. The
    remaining $60\%$ come from the redistribution of the open magnetic
  flux toward the equatorial plane (see \S\ref{sec:non-unif-puls}).}
  \label{fig:phivsalpha}
\end{figure}

A natural possibility is that the amount of open magnetic flux depends
on the inclination angle, $\Phi_{\rm open} = \Phi_{\rm open}(\alpha)$,
and this causes the dependence of the spindown luminosity on $\alpha$.
Fig.~\ref{fig:phivsalpha} shows that while this is indeed the
case to a degree, the  dependence $\Phi_{\rm open}(\alpha)$ is too
weak to fully account for the enhancement in spindown rate for oblique
pulsars. In fact, it accounts for about $40\%$ of the luminosity enhancement, as shown
with red dotted lines in Figs.~\ref{fig:Lvsalpha} and
\ref{fig:phivsalpha}. What is the cause of the rest of the
enhancement, i.e., the difference between the dotted red and solid
blue lines? 

\section{Non-uniformity of asymptotic pulsar wind and spindown rate}
\label{sec:non-unif-puls}

\begin{figure}
 \centering
 \includegraphics[width=1\columnwidth]{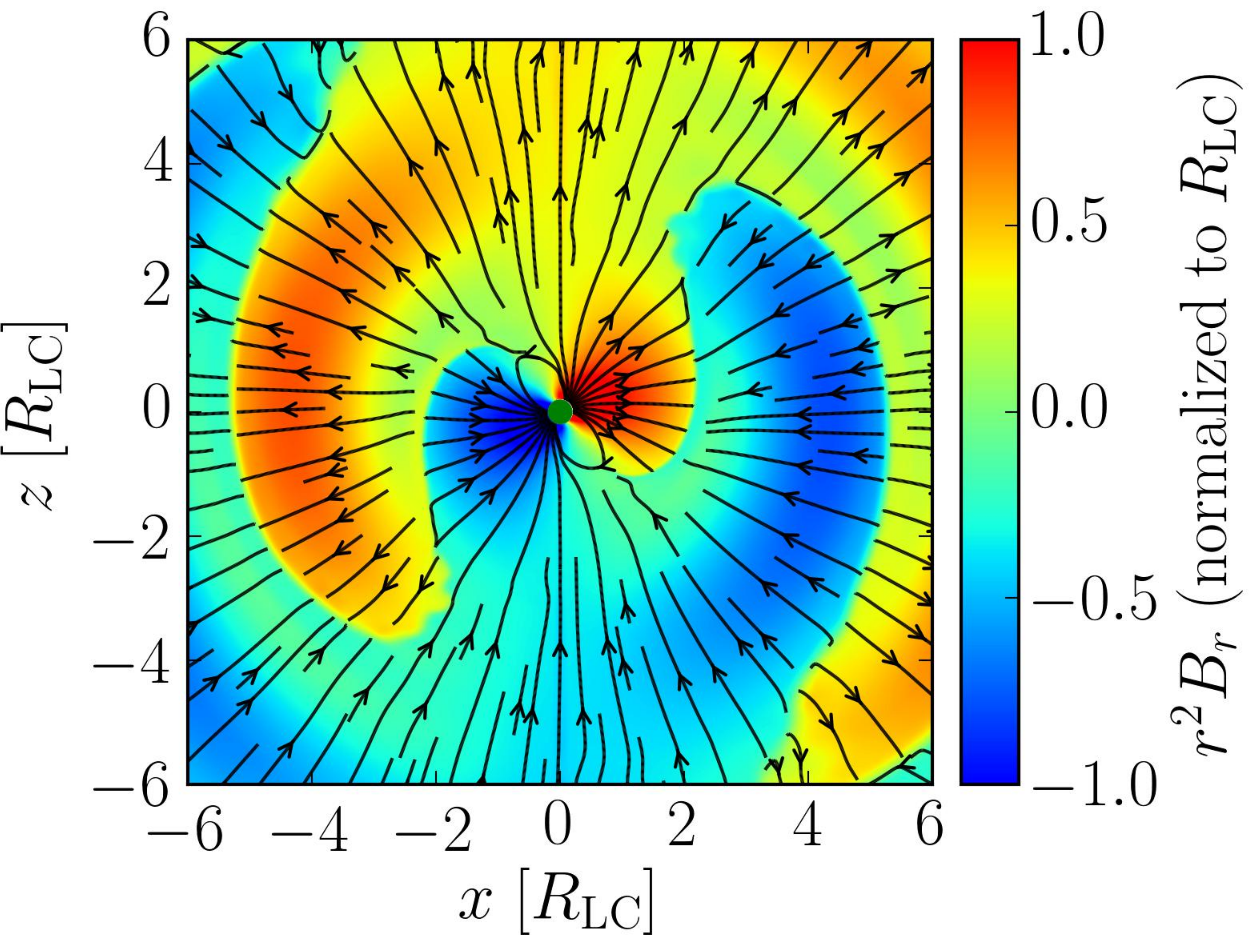}
 \caption{The same magnetospheric structure as
   Fig.~\ref{fig:pwind} but with colour showing scaled radial magnetic
   field, $r^2B_r$.  The $r^2-$prefactor eliminates the obvious
   radial trend: for a perfectly monopolar field, we would have
   $r^2B_r=$~const. The substantial deviations from constancy of
   $r^2B_r$ are a strong indication that the radial magnetic field in
   the pulsar magnetosphere is substantially non-uniform and deviates
   from the simple monopolar dependence, $B_r\propto1/r^2$.}
 \label{fig:pwindBr}
\end{figure}
As the colour variation in Fig.~\ref{fig:pwindBr} suggests, the product
 $|r^2B_r|$ is not constant in the simulated pulsar wind. 
However, the split-monopole model of the pulsar wind and the
associated spindown luminosity given by 
equation~\eqref{eq:Pmono} assume that $|r^2B_r|$ is a constant
everywhere, i.e., apart from $B_r$ changing sign at current sheet
crossings, $B_r$ is laterally-uniform (i.e., it has no dependence on $\theta$ and
$\varphi$) and follows a simple power-law dependence in
radius. According to Fig.~\ref{fig:pwindBr}, this approximation
is at best very rough: $B_r$ shows variations away from the expected
scaling, $1/r^2$, which indicates the presence of a substantial, order
unity non-uniformity in the pulsar wind.
Can these non-uniformities lead to deviations from the spindown rate given by
equation \eqref{eq:Pmono}? In section \S\ref{sec:veloc-freeze-cond} we will demonstrate that this is
indeed the case.

\subsection{Velocity and freeze-in condition in ideal MHD}
\label{sec:veloc-freeze-cond}

To describe the asymptotic structure of pulsar wind, we will make use
of the fact that the wind structure is
accurately described by a 2D distribution of just a single variable, $B_r$,
in $\theta{-}\varphi$ plane. This can be seen as follows. In ideal
MHD/force-free a pulsar magnetosphere rotates at 
the stellar angular frequency. Thus, the
velocity can be decomposed into motion along magnetic field lines with
velocity $v_\parallel$ and together with the magnetic field lines
corotating with the star at $v_\varphi = \Omega r\sin\theta$:
\begin{equation}
\vec v = v_\parallel \frac{\vec B}{\bigl|\vec B\bigr|} + \Omega r\sin\theta \vec e_\varphi.
\label{eq:vdecomposition}
\end{equation}
Due to the ideal MHD condition, the electric field is $\vec E = -(\vec
v/c)\times\vec B = - (\Omega r\sin\theta/c) B_r \vec e_\theta$. 

Asymptotically far from the pulsar, $r\sin\theta \gg R_{\rm LC}$, 
the magnetic field is toroidally-dominated, $B_\varphi
\gg B_r, B_\theta$ and moves outward at a drift velocity,
\begin{equation}
  \label{eq:vdr}
  v_{\rm dr}= \frac{c\bigl|\vec E\times \vec B\bigr|}{B^2} 
  \approx \frac{c E_\theta B_\varphi}{B_\varphi^2}.
\end{equation}
Since at such large distances
the wind moves relativistically fast, $v_{\rm dr} \approx c$, we have:
\begin{equation}
B_\varphi \approx -E_\theta = -(\Omega r\sin\theta/c) B_r.
\label{eq:EB}
\end{equation}
Using eq.~\eqref{eq:EB}, for a given distribution of $B_r$, we can
reconstruct the rest of the field components.

\subsection{Poynting flux}
\label{sec:poynting-flux}

Using eq.~\eqref{eq:EB}, we can write the
Poynting flux, which is carried by the outflowing toroidal magnetic
field, as
\begin{equation}
  \label{eq:spindown}
  S_r = \frac{c}{4\pi}\bigl(\vec E\times \vec B\bigr)_r \approx \frac{\Omega^2}{4\pi c}
   B_r^2r^2\sin^2\theta.
\end{equation}
Thus, the spindown luminosity is given by
\begin{equation}
  \label{eq:P}
  L = \iint S_r d\omega = 2\pi \int_0^\pi  \langle S_r\rangle_{\varphi} r^2 \sin\theta d\theta
  \approx \frac{\Omega^2}{2 c}\int_0^\pi \langle B_r^2\rangle_{\varphi} r^4 \sin^3\theta d\theta,
\end{equation}
where $\langle...\rangle_{\varphi}$ is the averaging over $\varphi$,
and we used the expression for the $\varphi-$average Poynting flux,
\begin{equation}
  \label{eq:dLdtheta}
  \avg{S_r}_\varphi 
  \equiv \biggl\langle\frac{\p L}{\p\omega}\biggr\rangle_\varphi 
  =\frac{\Omega^2}{4\pi c} \langle B_r^2\rangle_{\varphi} r^2 \sin^2\theta,
\end{equation}
that trivially follows from the definition of $S_r$, eq.~\eqref{eq:spindown}.
Both $L$ and its angular distribution $\avg{\p L/\p\omega}_\varphi$ 
are \emph{fully determined} by the surface distribution of a single
quantity, $B_r(\theta,\varphi)$, or, more precisely, by the
$\varphi$-average of its square, $\langle B_r^2\rangle_{\varphi}$.

If the magnetic field is split-monopolar, i.e.,
$|B_r| = B_0 (r_0/r)^{2}$ and
$B_0 = \Phi_{\rm open}/2\pi r_0^2$, eq.~\eqref{eq:P}
reduces to the standard split-monopolar expression,
eq.~\eqref{eq:Pmono}. 
Note, however, that the real pulsar wind is
non-uniform laterally, and this approximation does not capture this
nonuniformity.  However, even despite the non-uniformity, the pulsar wind
remains very nearly radial. This means that the radial scaling of
$B_r$ is a trivial power-law, $\propto r^{-2}$.

\subsection{Dipolar approximation for radial magnetic field}
\label{sec:dipol-appr-radi}

To make progress, we will approximate the angular distribution of
$B_r$ in the open magnetic flux region by
that of a dipole $\vec\mu$ inclined at an angle $\alpha$ in the
$x{-}z$, or $\varphi = 0$, plane. As we will
see below, this is an excellent approximation for
near-orthogonal rotators and a reasonable approximation for other
obliquities.  We obtain:
\begin{equation}
\label{eq:vacdipsol}
B_r = B_0(r/r_0)^{-2}\cos\theta_{\rm m},
\end{equation}
where $\theta_{\rm m}$ is magnetic colatitude, the angle away from the
magnetic axis,
\begin{equation}
\label{eq:thetamdef}
\theta_{\rm m} = \arccos(\sin\alpha\sin\theta\cos\varphi+\cos\theta\cos\alpha),
\end{equation}
 and 
\begin{equation}
\label{eq:B0dipole}
B_0 = \Phi_{\rm open}/\pi r_0^2.
\end{equation}
The angular distribution of $B_r$ due to eq.~\eqref{eq:vacdipsol} is clearly nonuniform: $B_r$ is strongest at the
magnetic poles ($\cos\theta_{\rm m} = \pm1$) and weakest along the magnetic
equator ($\cos\theta_{\rm m} = 0$). 

To find out how this nonuniformity affects the spindown, 
we first compute the angular distribution of $\varphi$-averaged $B_r^2$:
\begin{equation}
\langle B_r^2\rangle_\varphi = B_0^2(r/r_0)^{-4}(0.5\sin^2\alpha
\sin^2\theta + \cos^2 \theta \cos^2 \alpha).
\label{eq:Brsq}
\end{equation}
Now, by plugging eq.~\eqref{eq:Brsq} into eq.~\eqref{eq:P} and
integrating, we obtain:\footnote{Note that since in a dipole the open
  magnetic field depends on the pulsar period,
  $\Phi_{\rm open}\propto \Omega$, the dependence on the period is in
  $4$th power, $L_{\rm dipole} \propto \Omega^4$.  }
\begin{equation}
L_{\rm dipole}(\alpha) = \frac{\Omega^2\Phi_{\rm open}^2}{7.5\pi^2c} (1 + \sin^2 \alpha).
\label{eq:Pdip}
\end{equation} 
This is very close to the $\alpha$-dependence that is observed in
numerical simulations and shown in Fig.~\ref{fig:Lvsalpha} with the
solid blue line. 
Therefore,  the nonuniformity of $B_r$, via the magnetic flux concentration toward the
magnetic poles (i.e., around the direction of $\vec\mu$),
causes an \emph{enhancement in spindown losses} at high inclination
angles ($\alpha\sim90^\circ$).  

The qualitative behaviour given by eq.~\eqref{eq:Pdip} has a clear
physical interpretation. From eq.~\eqref{eq:spindown} we see that energy 
flux is minimum near the rotational poles ($\sin\theta\approx0$) and 
maximum near the rotational equator ($\sin\theta\approx1$). Hence,
the spindown losses are minimised when the maximum in $B_r$ is along
the rotational poles and does not contribute to the spindown
($\alpha = 0$). Conversely, the spindown losses are
maximised when the maximum in $B_r$ lies in the rotational equator and
contributes to the spindown most efficiently ($\alpha = 90^\circ$). 
In other words, the spindown losses are higher than in a pure monopole 
because of angular redistribution of magnetic flux from the polar to the 
equatorial regions.

From eqs.~\eqref{eq:vacdipsol} and
\eqref{eq:thetamdef}, the radial magnetic field strength of an
orthogonal rotator is:\footnote{This also coincides with $B_r$
  distribution of the full vacuum solution \citep{deutsch55}.}
\begin{equation}
B_r^{(2)}(r,\theta,\varphi)=B_0 (r/r_0)^{-2} \sin
\theta\cos\varphi.\quad ({\rm for\ orthogonal\ pulsars})
\label{eq:vacdipsol90}
\end{equation}
The azimuthally-average of \eqref{eq:vacdipsol90} is highly non-uniform
laterally (see also eq.~\ref{eq:Brsq}):
\begin{equation}
  \label{eq:Br2phiavganalytic}
  \avg{B_r^2}_\varphi = 0.5B_0^2 (r/r_0)^{-4}\sin^2\theta.
\end{equation}
This nonuniformity implies that the angular distribution of spindown
luminosity deviates 
from the standard expectation, $dL/d\omega\propto\sin^2\theta$,
eq.~\eqref{eq:dLdtheta}, and becomes,
\begin{equation}
  \label{eq:dL90domega}
\biggl\langle\frac{\p L_{\rm 90}}{\p\omega}\biggr\rangle_\varphi =
\frac{\Omega^2\Phi_{\rm open}^2}{8\pi^3 c r^2} \sin^4\theta, \quad
({\rm for\ orthogonal\ pulsars})
\end{equation}
where we used the eq.~\eqref{eq:B0dipole} to eliminate $B_0$ in favour
of $\Phi_{\rm open}$.

\section{Approximate Solution for Aligned and Orthogonal Force-free Pulsar Winds}
\label{sec:appr-solut-align}

\begin{figure}
\centering
\includegraphics[width=1\columnwidth]{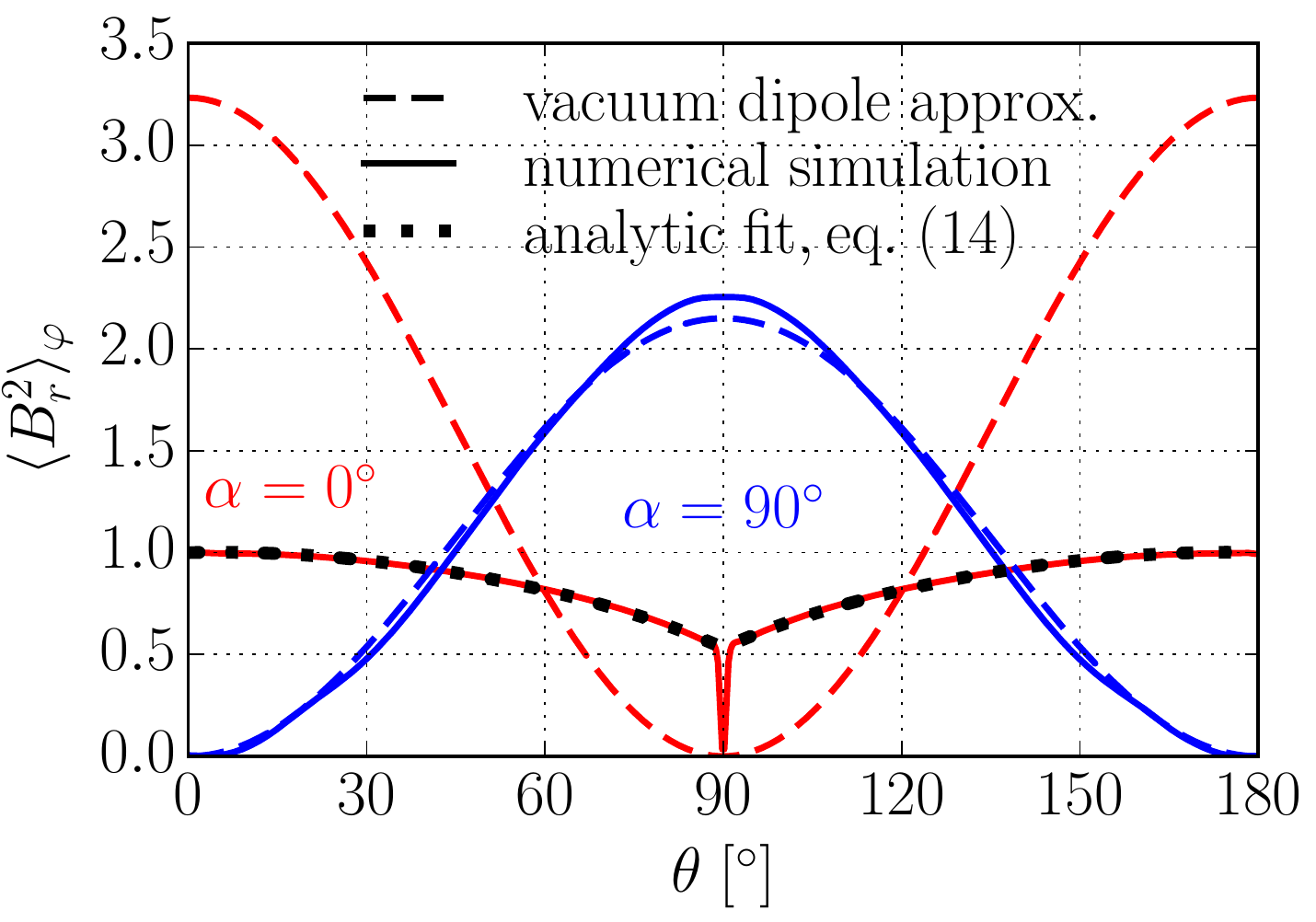}
\caption{The lateral distribution of azimuthally-averaged $B_r^2$
  (which determines the spindown rate, see eq.~\ref{eq:P}) due to (i) 
  numerical simulation shown with solid curves and due to (ii) the $B_r$
  distribution for vacuum dipole, eq.~\eqref{eq:vacdipsol}, with
  dashed curves. Red lines
  show the results for the aligned and solid lines for the orthogonal rotator.
  Both distributions are normalised to
  have the same magnetic flux. The vacuum dipole approximation
  accurately reproduces the lateral distribution of magnetic field for
  the orthogonal rotator, as shown with red curves.  However, it substantially
  over-predicts the polar field strength and under-predicts the equatorial
  field strength for the aligned rotator. The dotted black line shows an analytic fit
  \eqref{eq:mhdalignfit} for the force-free aligned rotator simulation
  results.}
\label{fig:Brsq_sim_vs_vacdip}
\end{figure}

\begin{figure*}
\centering
\includegraphics[width=\textwidth]{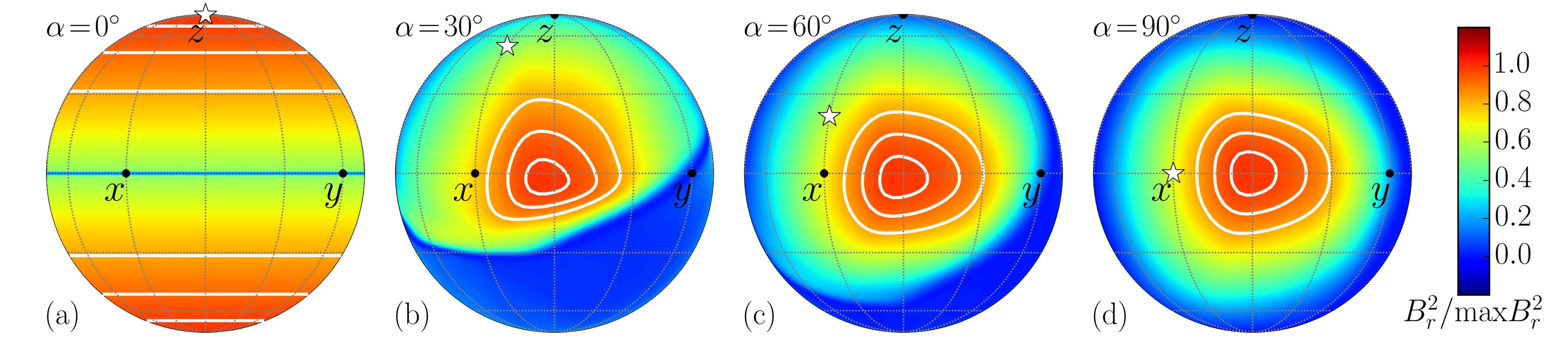}
\caption{Colour-coded surface distribution of $B_r^2$ in the numerical
  force-free solution (red shows high blue low values, see colour bar)
  for different inclination angles, at the surface of a sphere of
  radius $r = 6\Rlc$. The white contours indicate $91$st, $96$th, and
  $99$th percentile levels of $B_r$, which was Gaussian-smoothed with a standard
  deviation of $1.5$ cells. At all inclination angles the angular
  distribution of $B_r^2$ is clearly non-uniform. Even though for an
  aligned rotator, shown in {\bf panel (a)}, the maximum in $B_r^2$
  follows the polar field line (shown with the white asterisk), for
  other inclination angles {\bf (panels b-d)} the magnetic peak
  shifts: it moves laterally closer to equatorial plane and shifts in
  azimuth to a later phase by about
  $\varphi_{\rm max} \approx 30^\circ$ (see also Fig.~\ref{fig:thetas}).  Note that at nonzero
  inclination angles (panels b-d), the peak in the solution resembles
  that of the orthogonal rotator (panel d).}
\label{fig:Brsq_sim_map}
\end{figure*}

In Section \ref{sec:dipol-appr-radi}, we found that magnetic field
line bunching around the magnetic poles, such as in a vacuum dipole,
qualitatively reproduces the increased spindown rate of oblique force-free
and MHD
pulsars.  This is because most of the spindown emanates near the
equatorial plane, $\theta=\pi/2$, as indicated by the factor
$\sin^3\theta$ in eqs.~\eqref{eq:P} and \eqref{eq:dLdtheta}. Thus, for
a field line to contribute to spindown most efficiently, it needs to
be close to the equatorial plane. As we tilt the pulsar, its magnetic pole,
which is the region of enhanced magnetic field, approaches the
rotational equator. This leads to an enhancement in the spindown.

To find out whether this mechanism is at work in the simulated pulsar
magnetospheres, we compare the angular distributions of $B_r$ due
to a vacuum dipole solution, given by eq.~\eqref{eq:vacdipsol}, and in
a simulation of a force-free magnetosphere. Since we are
interested in the spindown rate, we will start by looking at an
azimuthal average, $\avg{B_r^2}_\varphi$, which
determines the spindown luminosity (see eq.~\ref{eq:P}).

Figure~\ref{fig:Brsq_sim_vs_vacdip} shows the comparison of the
vacuum dipole approximation given by eq.~\eqref{eq:vacdipsol} and
numerical force-free simulations for two limiting cases: aligned and
orthogonal rotators.  The vacuum dipole curves are normalised to have
the same open magnetic flux as in the numerical simulation (at the
same obliquity). 

From Fig.~\ref{fig:Brsq_sim_vs_vacdip}, we see that
the simulation result for the orthogonal rotator is well approximated by
eq.~(\ref{eq:Br2phiavganalytic}). Thus, the vacuum dipole provides an excellent
description of an orthogonal rotator wind.  We adopt
eq.~(\ref{eq:Br2phiavganalytic})  as the analytic description of the
magnetosphere of an orthogonal plasma-filled rotator. 

At the same time, Fig.~\ref{fig:Brsq_sim_vs_vacdip} shows that the vacuum dipole
expression gives a very poor description of the aligned rotator
solution: whereas in the simulation $B_r$ is non-zero in the
equatorial region, in the vacuum dipole field $B_r$ vanishes at the
equator. Since the total magnetic flux is the same, the polar field
strength in the vacuum model by a factor $\approx3$ exceeds that in the
simulation.  

For this reason, we adopt the following fit as the analytic approximation for the
aligned force-free solution:
\begin{multline}
  \label{eq:mhdalignfit}
  B_r^{(1)}(\theta) = B_0(r/r_0)^{-2}
  \bigl[1+0.02\sin\theta+0.22(\cos\theta-1) \\
  -0.07(\cos\theta-1)^4\bigr]\times{\rm sign}\cos\theta.
\end{multline}
This fit, shown in Fig.~\ref{fig:Brsq_sim_vs_vacdip} with the black
dotted line, provides an adequate quantitative description of the
aligned force-free solution (at a representative distance $r = 6\Rlc$
by which the solution has approximately settled to an asymptotic
distribution). We discuss the lateral structure of this distribution in
Appendix~\ref{sec:appendix}.

\section{Pulsar wind structure in magnetic coordinates}
\label{sec:puls-wind-struct}

In \S\ref{sec:non-unif-puls}, we obtained analytic approximations,
eqs.~\eqref{eq:vacdipsol90} and~\eqref{eq:mhdalignfit}, for
$B_r(\theta,\varphi)$ -- and thus an approximate description of
the pulsar wind -- for the two limiting cases of aligned and
orthogonal rotators. Our goal is to obtain the description for
pulsar wind at all inclination angles. Could we somehow combine these two expressions in
order to get a general solution? With
this in mind, let us a look at the structure of the force-free pulsar
winds simulations at various inclinations.

Fig.~\ref{fig:Brsq_sim_map} shows the map of $B_r^2$ in the numerical
force-free solution of a pulsar wind  on the surface of
a sphere of radius $r_0 = 6\Rlc$. By this distance, the structure of
the wind has frozen out and does not change at larger distances. 
Panels (a)--(d) illustrate
solutions for different inclination angles, $\alpha = 0, 30, 60$, and
$90^\circ$, respectively, in two ways: (i) via a colour map (see the
colour bar) and (ii) with white contours that indicate $91$, $96$, and
$99$ percentile levels of $\max |B_r|$.  

For each inclination angle,
$B_r^2$ exhibits a well-defined peak. The distribution of
$B_r$ for the aligned rotator, shown in
Fig.~\ref{fig:Brsq_sim_map}(a), peaks at the rotational $z-$axis, as
consistent with Fig.~\ref{fig:Brsq_sim_vs_vacdip}. 

\begin{figure}
\centering
\includegraphics[width=\columnwidth]{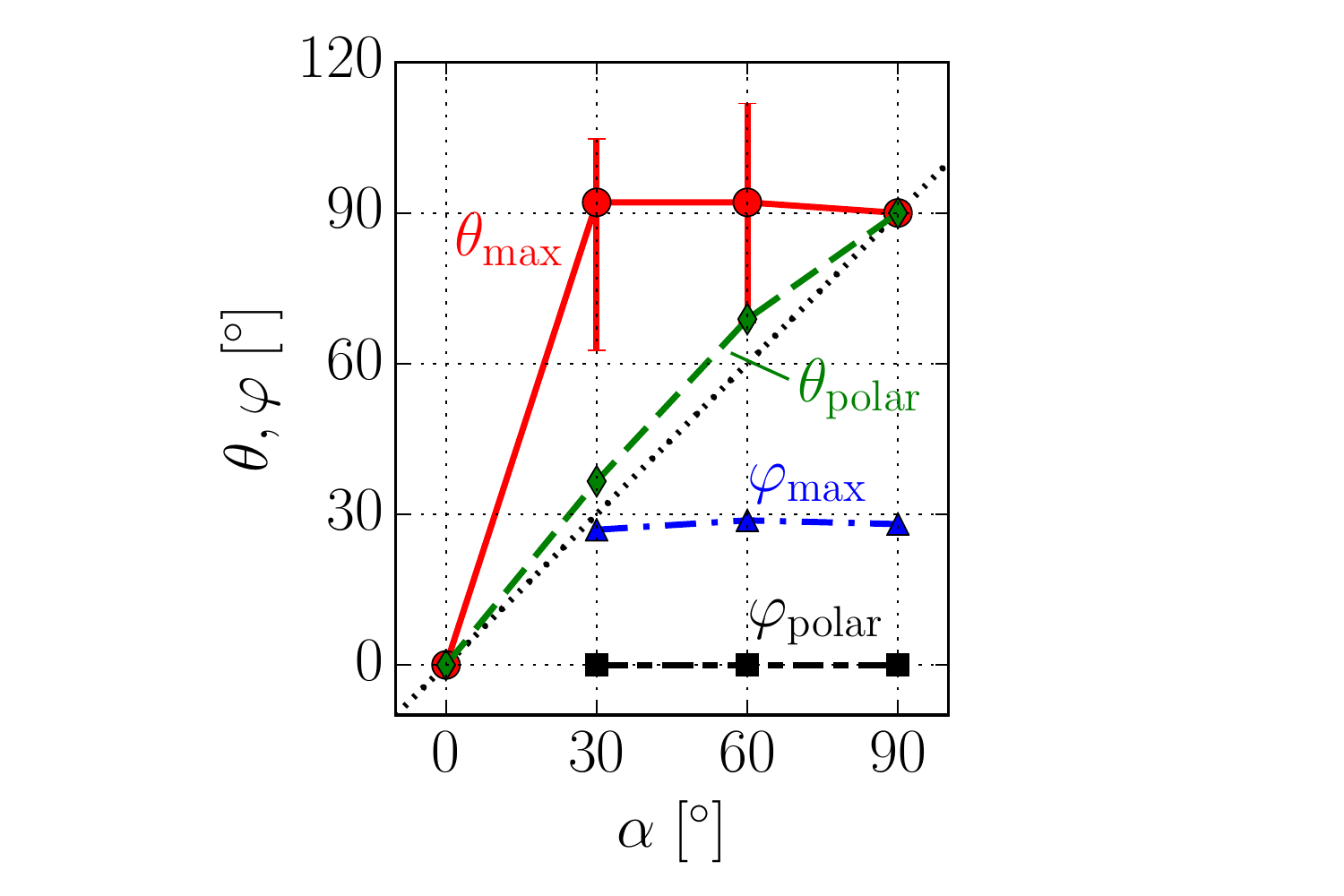}
\caption{Positions of the polar field line and maximum in $B_r$ are
  plotted with various lines vs the inclination angle, $\alpha$, as
  measured at the sphere of radius $r = 6\Rlc$.
  Shown are the polar coordinates of the maximum in magnetic field,
  ($\theta_{\rm max}$, $\varphi_{\rm max}$), with error bars indicating
  the range 
  spanned by the outermost white contour in
  Fig.~\ref{fig:Brsq_sim_map}.  We also plot the coordinates of the polar
  field line ($\theta_{\rm polar}$, $\varphi_{\rm polar}$), as labeled
  in the figure.  The maximum of the field, shown with blue connected
  triangles, is located near the equatorial plane and outruns the
  polar field line by $\gtrsim30^\circ$ for
  $\alpha\sim30{-}60^\circ$. 
  The polar angle of the polar
  field line, which is shown with the green connected diamonds,
  approximately follows the inclination angle, i.e., the polar field
  line, despite being wrapped around in the azimuthal direction,
  maintains its polar angle. By definition we use the intersection
  point with the polar field line to set the zero point in the
  azimuth, i.e., we set $\varphi_{\rm polar} = 0$.}
\label{fig:thetas}
\end{figure}

\begin{figure}
\begin{center}
    \includegraphics[width=1\columnwidth]{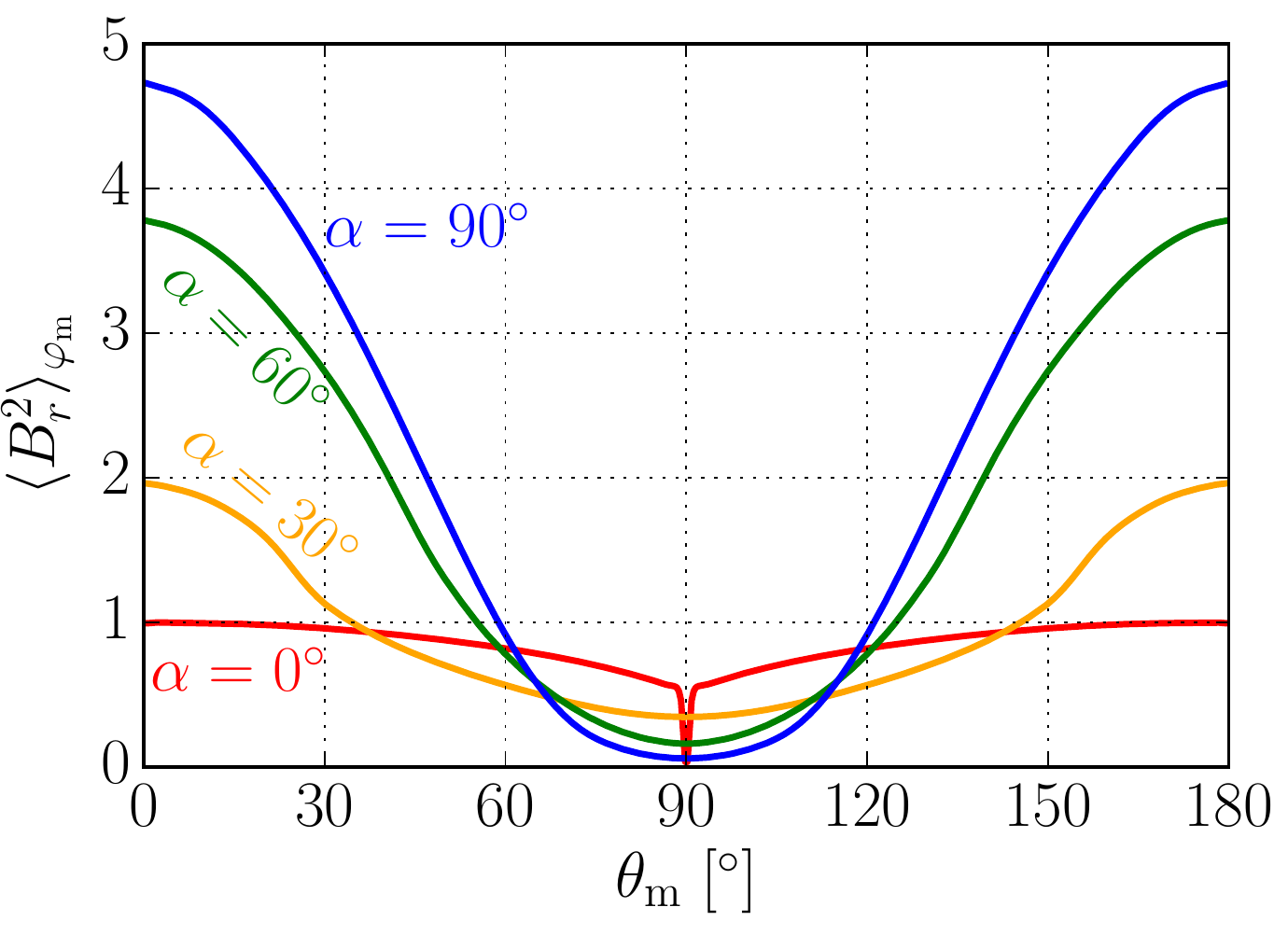}
  \end{center}
\caption{Angular distribution of the $\langle
  B_r^2\rangle_{\varphi_m}$ of the numerical force-free solution in
  \emph{magnetic coordinates} morphs from that of an aligned rotator at
  $\alpha = 0^\circ$ (red line) to that of an orthogonal rotator at
  $\alpha = 90^\circ$ (blue line). The in-between distributions, for
  intermediate inclination angles, $0^\circ < \alpha < 90^\circ$
  resemble neither of these two limiting solutions, making it
  difficult to construct an analytic solution for those obliquities.}  
\label{fig:brsqphimavg}
\end{figure}

\begin{figure*}
\centering
\includegraphics[width=\textwidth]{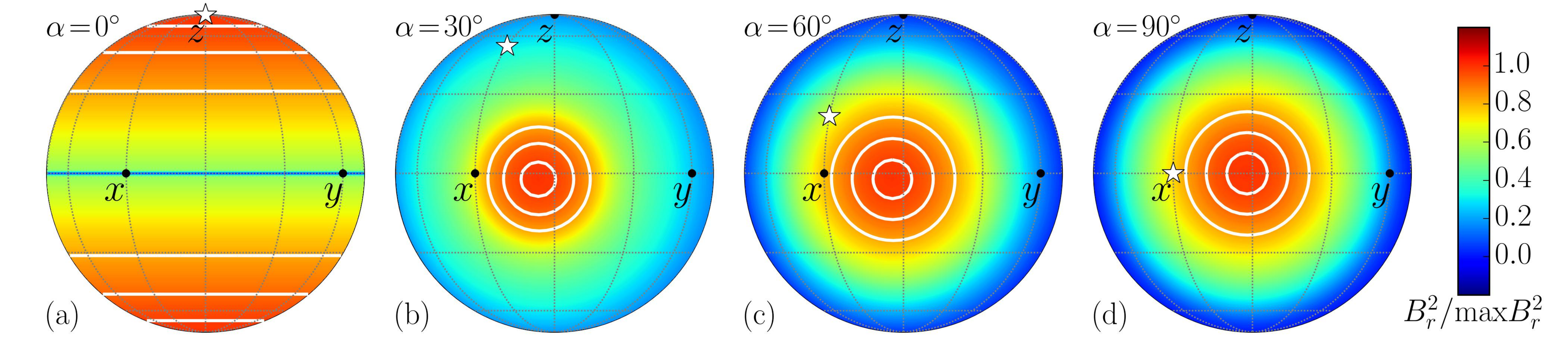}
\caption{Colour-coded surface distribution of $B_r^2$ in the numerical
  force-free solution that was averaged out around the magnetic peak.  For
    inclinations $\alpha\gtrsim60^\circ$, this axial averaging
    approach gives a good approximation of solution structure around
    the peak. However, the averaging washes out the current sheet,
    which is a source of magnetospheric dissipation and potential
    origin of gamma-ray emission. See Fig.~\ref{fig:Brsq_sim_map} for
    more details.}
\label{fig:Brsq_phimavg_map}
\end{figure*}

\begin{figure}
\begin{center}
    \includegraphics[width=1\columnwidth]{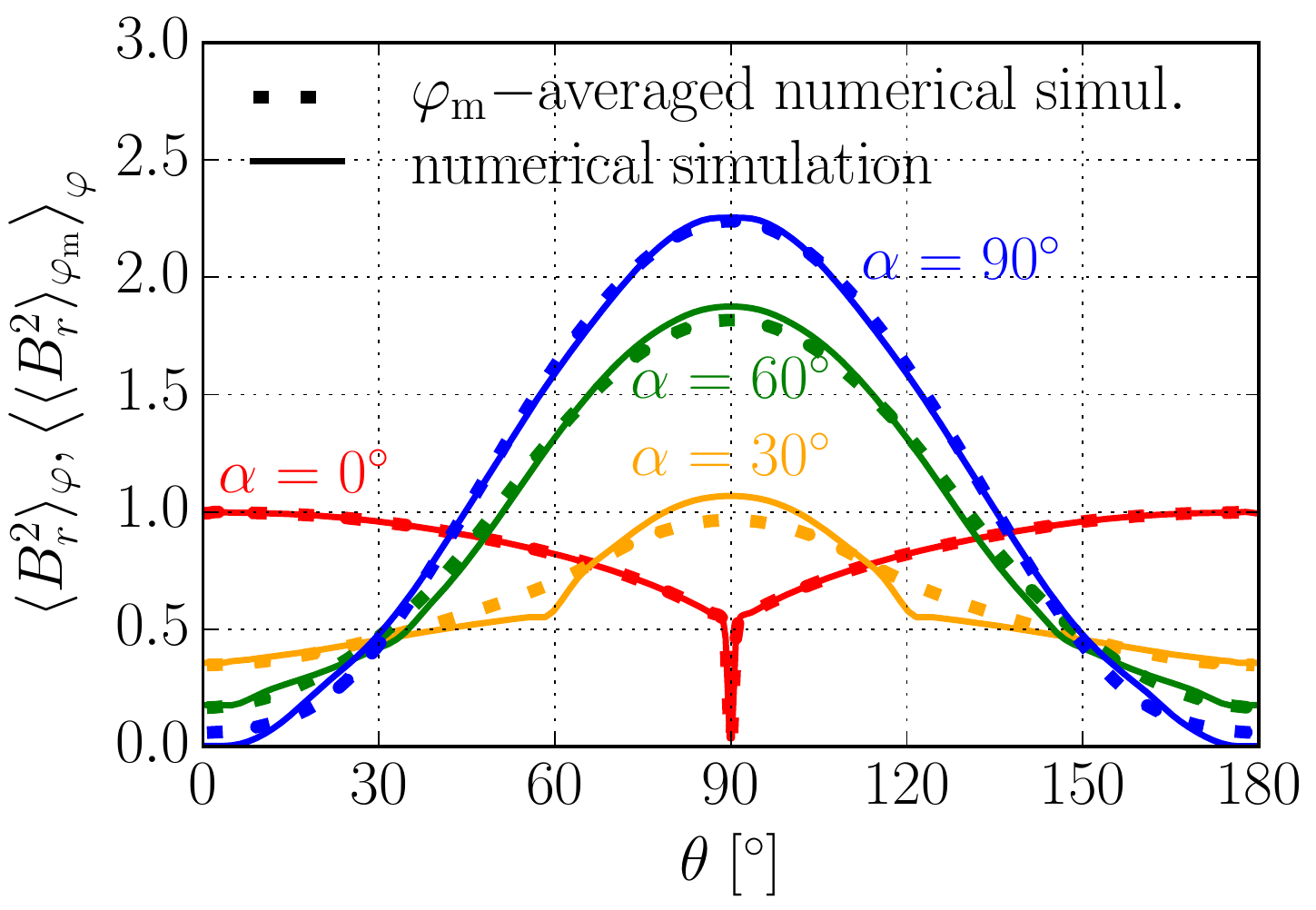}
  \end{center}
  \caption{Angular distribution of the $\langle B_r^2\rangle_\varphi$ in
    force-free simulations (solid lines) and same numerical simulation
    result but additionally $\varphi_{\rm m}-$averaged around the
    magnetic peak, $\bigl\langle \langle
    B_r^2\rangle_{\varphi_{\rm m}}\bigr\rangle_{\rm \varphi}$ (dotted
    lines) for different obliquities. The
    angle-averaging around the magnetic pole gives a good
    approximation for $\alpha = 0^\circ$ and $\alpha\gtrsim60^\circ$ but washes out the
    non-axisymmetric structure around the pole at intermediate
    inclinations, which manifests itself as the
    smoothing out of the equatorial peak in the dotted orange $\bigl\langle \langle
    B_r^2\rangle_{\varphi_{\rm m}}\bigr\rangle_{\rm \varphi}$ for
    $\alpha = 30^\circ$.}  
\label{fig:bsqvstheta}
\end{figure}

How does the situation change for an oblique rotator?  We show the
solution for a $30$--degree inclined rotator in
Fig.~\ref{fig:Brsq_sim_map}(b). The white star shows the point at
which the polar field line pierces our sphere of $r = 6\Rlc$.
We denote this position as ($\theta_{\rm polar}, \varphi_{\rm polar}$)
and use it to
define the zero meridian, and with it our $x{-}z$
plane. Thus, by definition we set $\varphi_{\rm polar} = 0$. 
Figs.~\ref{fig:Brsq_sim_map} and \ref{fig:thetas} show that
$\theta_{\rm polar}$ exceeds the inclination angle by only a few
degrees: this means that as the polar field line propagates out and
winds up around the rotational axis, it does so mostly along a
cone of a constant opening angle.

However, this does not mean that the entire magnetosphere tilts
rigidly and retains its lateral structure. If it did so, we would
expect the peak in $B_r$ to shift by the inclination angle, i.e. we
would expect to have $\theta_{\rm max}=\alpha$.
Figures~\ref{fig:Brsq_sim_map}(b) and \ref{fig:thetas} show that this
does not happen: for a stellar tilt of $\alpha = 30^\circ$, the peak
tilts by three times as much, $\theta_{\rm max} \approx 90^\circ$.
Figs.~\ref{fig:Brsq_sim_map}(c) and \ref{fig:thetas} show that this
trend persists for higher inclination angles: e.g., for
$\alpha = 60^\circ$, we also find $\theta_{\rm max} \approx 90^\circ$. The polar
field line and the peak in $B_r$ only catch up with each other for an
orthogonal rotator, $\alpha=\theta_{\rm max}=90^\circ$.  Therefore, a
tilted magnetosphere rearranges itself in a non-trivial way, with
different parts moving relative to each other and causing the maximum
of radial magnetic field strength to shift toward the equatorial plane
by more than we would naively expect.

Note that the above picture applies at a rather large distance from
the stellar surface, $r = 6\Rlc$. At much smaller distances, e.g., at
the surface of the star, the magnetic peak and the polar field line
coincide. They smoothly diverge from each other with increasing
distance from the stellar surface. The solution shown in
Fig.~\ref{fig:Brsq_sim_map} qualitatively persists out to much larger
distances; asymptotically far away from the star, however, it
undergoes small quantitative changes. We expect that at larger
distances, $r>6\Rlc$, the solution does not have much opportunity to
rearrange laterally: this is because the Lorentz factor of the wind
scales as $\gamma \approx R/\Rlc\gg1$, and hence we expect that the
solution might only be able to change on the small angular scales,
$\delta\theta \sim 1/\gamma \sim \Rlc/R \lesssim 1/6 \approx 10^\circ$ for
$R > 6\Rlc$.

Despite this magnetospheric rearrangement, the surface distribution of
$B_r^2(r_0,\theta,\varphi)$ appears to maintain a rather high degree of
symmetry around the magnetic peak,
$(\theta_{\rm max}, \varphi_{\rm max})$. Are the deviations from
axisymmetry around the peak important for the angular distribution of
emergent luminosity?  To test this, we plot in
Fig.~\ref{fig:brsqphimavg} the magnetic field distribution vs the
magnetic co-latitude $\theta_{\rm m}$ (which is counted off from the
magnetic peak), averaged over the magnetic azimuth $\varphi_{\rm m}$
(relative to the magnetic peak), $\avg{B_r^2}_{\rm \varphi_{\rm m}}$.
We see that as the inclination angle increases, the profile of
$\avg{B_r^2}_{\rm \varphi_{\rm m}}$ smoothly transitions from that of
an aligned pulsar (red line, $\alpha = 0^\circ$) to that of a vacuum
dipole, $B_r^2 \propto \cos^2\theta_{\rm m}$ (blue line,
$\avg{B_r^2}_{\rm \varphi_{\rm m}}$). This is another indication that
the vacuum dipole magnetic field distribution provides a good
description of the orthogonal pulsar.

\begin{figure*}
\centering
\includegraphics[width=\textwidth]{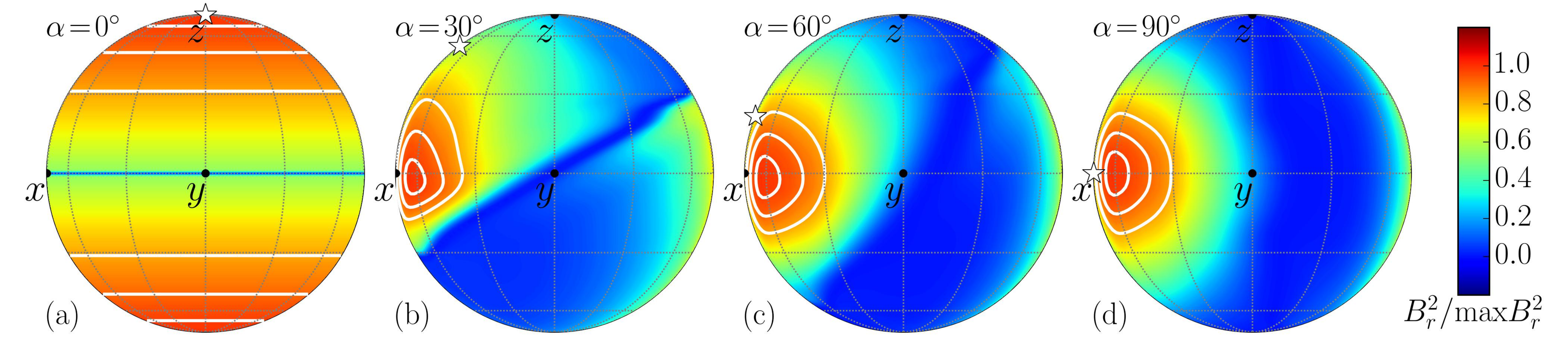}
\caption{Colour-coded surface distribution of $B_r^2$ in the
  numerical force-free solution at $r = 6\Rlc$ viewed down the
  $y-$axis. This is similar to Fig.~\ref{fig:Brsq_sim_map} where the
  same solution is viewed down the magnetic peak. 
  For all inclination angles, except $\alpha = 90^\circ$, the solution
  shows a well-defined current sheet whose orientation is well-described by the
  split-monopole model shown in Fig.~\ref{fig:Brsq_monoyaxis_map}.}
\label{fig:Brsq_numericalyaxis_map}
\end{figure*}
\begin{figure*}
\centering
\includegraphics[width=\textwidth]{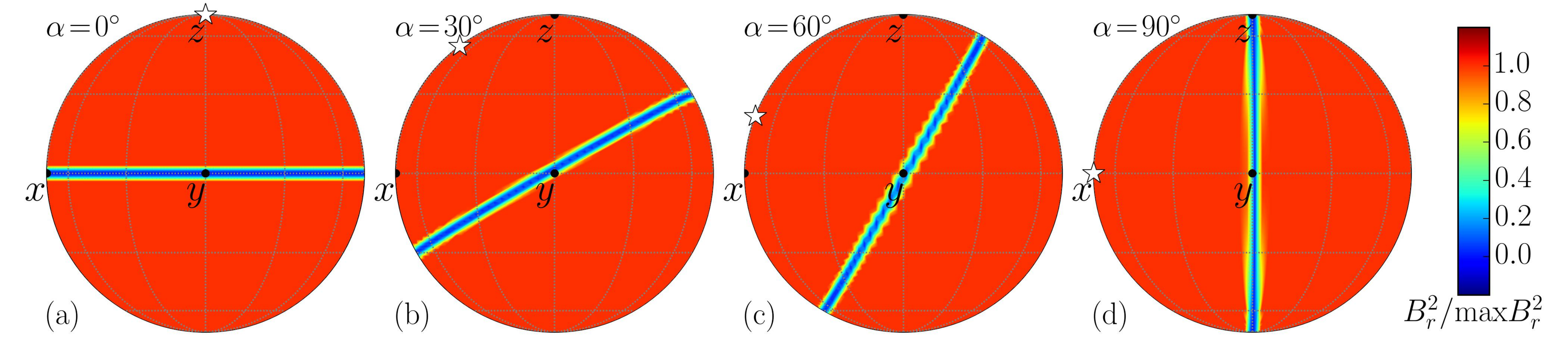}
\caption{Colour-coded surface distribution of $B_r^2$ in the
  split-monopole solution \citep{Bogovalov}. The current sheet, in which the radial
  magnetic field vanishes, describes the orientation of the current
  sheet in the   numerical force-free solutions shown in Fig.~\ref{fig:Brsq_sim_map}.}
\label{fig:Brsq_monoyaxis_map}
\end{figure*}

Fig.~\ref{fig:Brsq_phimavg_map} shows that after averaging out in
$\varphi_{\rm m}$, the solution visually appears very similar to the
full force-free simulation.  Will this approximation give a similar
spindown compared to the simulation?  To test this, in Fig.
Fig.~\ref{fig:bsqvstheta} we show the resulting distribution of
$\bigl\langle\avg{B_r^2}_{\varphi_{\rm m}}\bigr\rangle_{\varphi}$ with
dotted lines and the results of the full force-free simulation with
solid lines. The agreement is excellent at all inclination angles: the
differences are generally small and are noticeable mostly at small
values of $\alpha$. For instance, at $\alpha = 30^\circ$, the
approximation smoothes out the otherwise well-defined equatorial peak
in $\langle B_r^2\rangle_\varphi$.

Since most of the spindown emerges near the equatorial plane --- and
these regions are rather well-described by this approximation --- we
expect that it provides a decent description of pulsar spindown.
However, there are two difficulties in using this description in
practice. First, it requires knowing the shapes plotted in
Fig.~\ref{fig:brsqphimavg} for all inclination angles. Second, it
ignores the presence of the equatorial current sheet: in fact, the
averaging in $\varphi_{\rm m}$ ``washes out'' the jump in the magnetic
field across the current sheet, which is clearly seen as a jump in
$B_r$ in the bottom right of Fig.~\ref{fig:Brsq_sim_map}(b),(c). Since
the current sheet could be an important location at which dissipation
takes place and magnetospheric radio and gamma-ray emission is
produced, we will try to find a solution that
properly describes the current sheet.

\section{Semi-analytical solution to the pulsar wind}
\label{sec:semi-analyt-solut}

The current sheet structure in the force-free solutions is seen more clearly
in Fig.~\ref{fig:Brsq_numericalyaxis_map}, which shows the view of the
solution down the $y-$axis. The current sheet is clearly seen as the
sharp stripe across the solution. Apart from a small-scale deformations,
the current sheet appears to lie in a single plane inclined relative
to the $x{-}y$ plane.
 
A common
description of the magnetospheric current sheet is due to
\hbox{\citet{Bogovalov}} split-monopole model. This approach gives a 3D,
analytic description of the magnetospheric current sheet location
under the assumption that the sheet is ``painted'' on the
magnetosphere and is passively advected outward at the speed of light,
as shown in Fig.~\ref{fig:Brsq_monoyaxis_map}.  This model is in
rather good agreement with the numerical
simulation shown in Fig.~\ref{fig:Brsq_numericalyaxis_map} and
captures both the inclination and the phase of the current
sheet correctly.  However, the split-monopole model assumes a piece-wise constant
$B_r$ outside of the current sheet, and this is clearly not the case in the
numerical simulations.

\begin{figure*}
\centering
{\large Numerical force-free solution at $r = 6\Rlc$:}
\includegraphics[width=\textwidth]{Brsq_spheremap_numericalyaxis.pdf}
{\large Analytical approximation:}
\includegraphics[width=\textwidth]{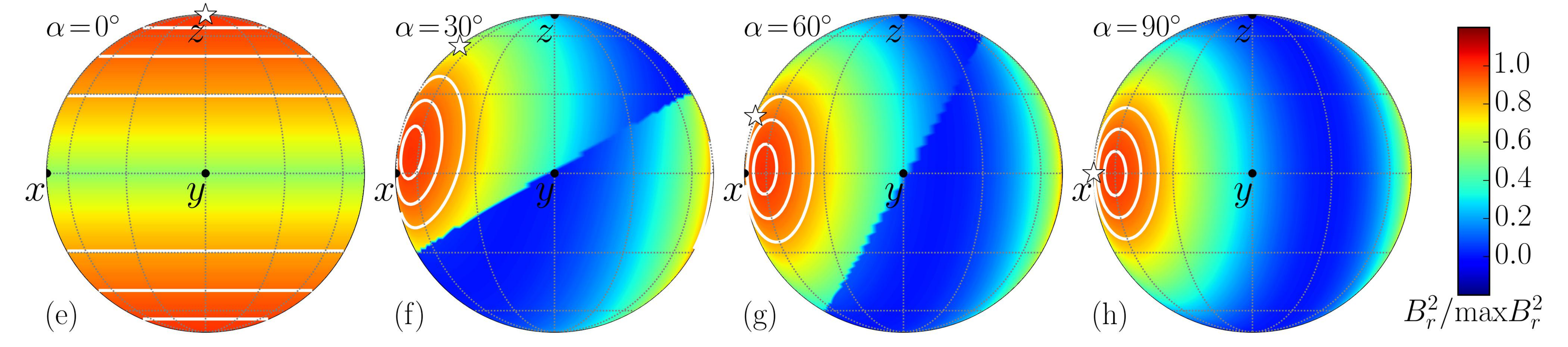}
\caption{Colour-coded surface distribution of $B_r^2$ in the numerical
  (panels a-d) and analytical (panels e-h) solutions. For convenience
  of comparison, the top row of panels duplicates the numerical
  solution shown in Fig.~\ref{fig:Brsq_numericalyaxis_map}. The
  analytical solution, shown in the bottom row of panels, clearly
  reproduces the main features of the numerical solution for all
  inclination angles. Firstly, the position of the current sheet (the
  jump in the magnitude of $B_r^2$) are in good agreement. Secondly,
  the position and size of the peaks in $B_r^2$ are consistent with
  each other to within the middle ($96$th percentile) contour
  level. Thus, the analytical approximation provides an excellent
  description of the numerical solution.}
\label{fig:Brsq_analyticalyaxis_map}
\end{figure*}

\begin{figure*}
\centering
{\large Numerical force-free solution at $r = 6\Rlc$:}
\includegraphics[width=\textwidth]{Brsq_spheremap_numerical.pdf}
{\large Analytical approximation:}
\includegraphics[width=\textwidth]{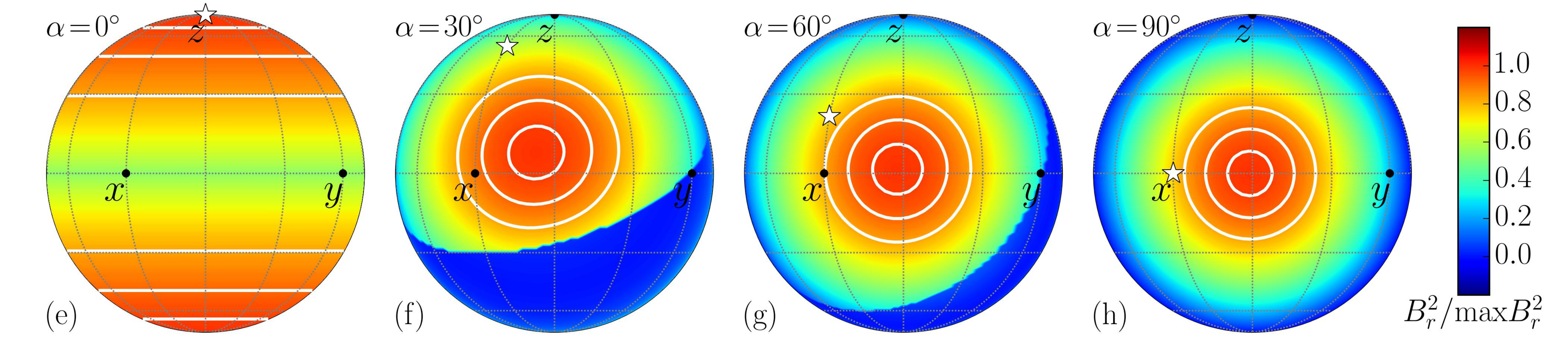}
\caption{Same as in Fig.~\ref{fig:Brsq_analyticalyaxis_map} but
  centred at the position of the magnetic peak. The analytic
  model, shown in panels (a)--(d) provides an accurate description of
  the numerical solution at all inclination angles. }
\label{fig:Brsq_analytical_map}
\end{figure*}

Our goal is to construct an approximate analytic description of the 3D
pulsar wind that would correctly reproduce both the location of the current
sheet and the smooth variation of the magnetic
field everywhere else.  In particular, the solution needs to capture
the gradual decrease of the jump in magnetic field strength across the
current sheet as the inclination angle increases from $\alpha =
30^\circ$ to $90^\circ$, as seen in
Fig.~\ref{fig:Brsq_numericalyaxis_map}(b)--(d). This might seem like an intractable problem.
However, as we discussed in \S\ref{sec:dependence-spin-down}, the
situation is substantially simplified by the fact that to describe the
entire wind solution it suffices to specify just one function,
$B_r = B_r^0(\theta,\varphi)$, at a sufficiently large radius
$r_0\gg\Rlc$ at which the lateral structure of the solution ``freezes out''. After this, the rest of the solution can be unambiguously
reconstructed.
As in the split-monopole model pulsar wind \citep{Bogovalov}, this
distribution rotates at an angular frequency $\Omega$ of the star in the
$\varphi-$direction and ``flies out'' at essentially the speed of light in the
$r-$direction. The full time-dependent solution combines both of these
effects:
\begin{equation}
B_r = B_r^0(\alpha,\theta,\varphi+\Delta\varphi) \left(\frac{r}{r_0}\right)^{-2},
\label{eq:Brfullsol}
\end{equation}
where the phase shift is
$\Delta\varphi=\Omega\cdot[(r-r_0)/c-(t-t_0)]$, and we approximate
$B_\theta = 0$, as the wind is mostly radial at large distances (see
Figs.~\ref{fig:pwind} and \ref{fig:pwindBr}). Then,
the toroidal magnetic field is given by the winding of this radial
field eq.~\eqref{eq:EB} (see also \citealt{2006ApJ...647L.119G}).
Within the approximation $B_\varphi=-E_\theta$ that we will adopt, the
Lorentz factor of the flow is
\begin{equation}
\label{eq:gamma}
\gamma = \frac{B}{(B^2-E^2)^{1/2}} = \frac{B}{B_r} 
= (1 + \Omega^2 r^2 \sin^2\theta)^{1/2} \approx \Omega r\sin\theta/c
\end{equation}
\citep{tch08}.  In the simulations, we find that
\hbox{$B_\varphi^2 - E_\theta^2$} $\lesssim B_r^2\ll B_\varphi^2,E_\theta^2$, so the Lorentz
factor could be higher or lower by a factor $\lesssim2$ than the estimate
\eqref{eq:gamma}. For simplicity, we will neglect this effect.

Our goal is to construct an appropriate 2D function,
$B_r^0(\theta,\varphi)$, that would satisfy the above
requirements. As we discussed in \S\ref{sec:puls-wind-struct}, the
structure of the pulsar wind is relatively simple in the two limiting
cases: 
\begin{enumerate}
\item \label{item.i} aligned rotator ($\alpha = 0^\circ$). While no
  analytic solution is known, we can write down a simple analytic
  fitting function, eq.~\eqref{eq:mhdalignfit}, for the numerical
  solution. We will denote this solution, rotated by an angle $\alpha$
  around the $y-$axis, via
  $B_r^{(1)}(r,\alpha,\theta,\varphi)\equiv B_r^{(1)}\left[r,\theta_{\rm
      m}(\alpha,\theta,\varphi),\varphi\right]$,
  where $B_r^{(1)}(r,\theta,\varphi)$ is given by
  eq.~\eqref{eq:mhdalignfit} and
  $\theta_{\rm m}(\alpha,\theta,\varphi)$ by eq~\eqref{eq:thetamdef}.
\item \label{item.ii} orthogonal rotator ($\alpha = 90^\circ$), whose
  $B_r$ distribution is well-described by  eq.~\eqref{eq:vacdipsol90}. 
  We will denote this solution as $B_r^{(2)}(r,\theta,\varphi)$.
\end{enumerate}

Our goal is to describe what happens in between these two limiting
cases. 
We
established that the position of the current sheet is well-described
by the split-monopole model, i.e, the current sheet lies in the plane
that passes through the $y-$axis and makes an angle $\alpha$, i.e, the
inclination angle, with the $x{-}y$ plane. We can incorporate this
feature by tilting the solution for the aligned rotator 
by the inclination angle, $\alpha$, as described in item \ref{item.i}. 

This is clearly not the whole story since this would lead to the
magnetic peak located at $\theta_{\rm max} = \alpha$ in the $x{-}z$
plane. However, we saw that the position of the magnetic peak is
systematically shifted toward the equatorial plane, i.e.,
$\theta_{\rm max} > \alpha$. Further, the strength of the jump across
the current sheet progressively decreases with increasing
$\alpha$. Eventually, for $\alpha = 90^\circ$, the solution becomes
the orthogonal solution [see item \ref{item.ii}] without any current sheet.

On a phenomenological level, this suggests that it would be
advantageous to mix the tilted solution \ref{item.i} with the
orthogonal one \ref{item.ii}.  This might work because the magnetic
peak of the orthogonal rotator is always located at the midplane and
will bias the sum to peak toward the midplane:
\begin{equation}
  \label{eq:Br0}
  B_r^0(\alpha,\theta,\varphi) = \left[w_1(\alpha)B_r^{(1)}(\alpha,\theta,\varphi) + w_2(\alpha)B_r^{(2)}(r,\theta,\varphi-\varphi_{\rm max})\right]\psi(\alpha),
\end{equation}
where $\psi(\alpha)$ describes the variation of total open flux vs the
inclination angle, as seen in Fig.~\ref{fig:phivsalpha}. Here
$\varphi_{\rm max}=30^\circ$ is the approximate azimuth of the
magnetic peak, as seen in Fig.~\ref{fig:thetas}.
We can determine the weights $w_1$, $w_2$, and $\psi$ in the following way.
Since the orthogonal solution \ref{item.ii} vanishes at the rotational
pole, i.e., $B_r^{(1)}(\theta=0,\varphi) =B_r^{(1)}(\theta=\pi,\varphi) = 0$, we choose the value of
$w_2$ by requiring that we get the polar value of $B_r$ from the
simulation. Next, we choose $w_2$ such that the weighted sum of the two
solutions in the square brackets in eq.~\eqref{eq:Br0} gives us an inclination-independent value of the total magnetic flux. Then, we choose $\psi$ to reproduce the dependence of the open magnetic flux on the
inclination angle, as seen in the simulations and in
Fig.~\ref{fig:phivsalpha}. This way we find the weights of the aligned
and orthogonal components of the full solution, respectively,
\begin{align}
\label{eq:w1}
w_1(\alpha) &= \left|1-2\alpha/\pi\right|,\\
\label{eq:w2}
w_2(\alpha) &= 1+0.17\left|\sin2\alpha\right|-w_1,\\
\intertext{and the overall normalisation factor that accounts for the variation
of the open magnetic flux with the inclination angle,}
\label{eq:psi}
\psi(\alpha)&=1+0.2\sin^2\alpha.
\end{align}
We have found that eq.~\eqref{eq:Br0} with the weights given by
eqs.~\eqref{eq:w1}--\eqref{eq:psi} provide a good description of the numerical
simulation results. We discuss the resulting two-component solution below.

Figs.~\ref{fig:Brsq_analyticalyaxis_map} and
\ref{fig:Brsq_analytical_map} show the comparison of such a
two-component approximation to the numerical solution from two viewing
angles: one along the $y-$axis and the other along the magnetic peak,
respectively.  Fig.~\ref{fig:Brsq_analyticalyaxis_map} shows that our
analytic description accurately reproduces not only the position of
the magnetospheric current sheet but also the magnitude of the jump in
$B_r^2$ across the magnetospheric current sheet. This highlights an
important difference from the \cite{Bogovalov} split-monopole solution,
in which $B_r$ changes sign across the current sheet but maintains the
same absolute magnitude, i.e., there is no jump in $B_r^2$ across the
current sheet (see
Fig.~\ref{fig:Brsq_monoyaxis_map}). 

Fig.~\ref{fig:Brsq_analytical_map} shows that our analytic approximation
is in excellent agreement with the numerical simulation outside of
the current sheet region. In particular, for all inclination angles we
considered, the analytic description captures the angular size of the
magnetic peak and its location (to within $4$\% in
$B_r^2$).

\begin{figure}
\begin{center}
    \includegraphics[width=1\columnwidth]{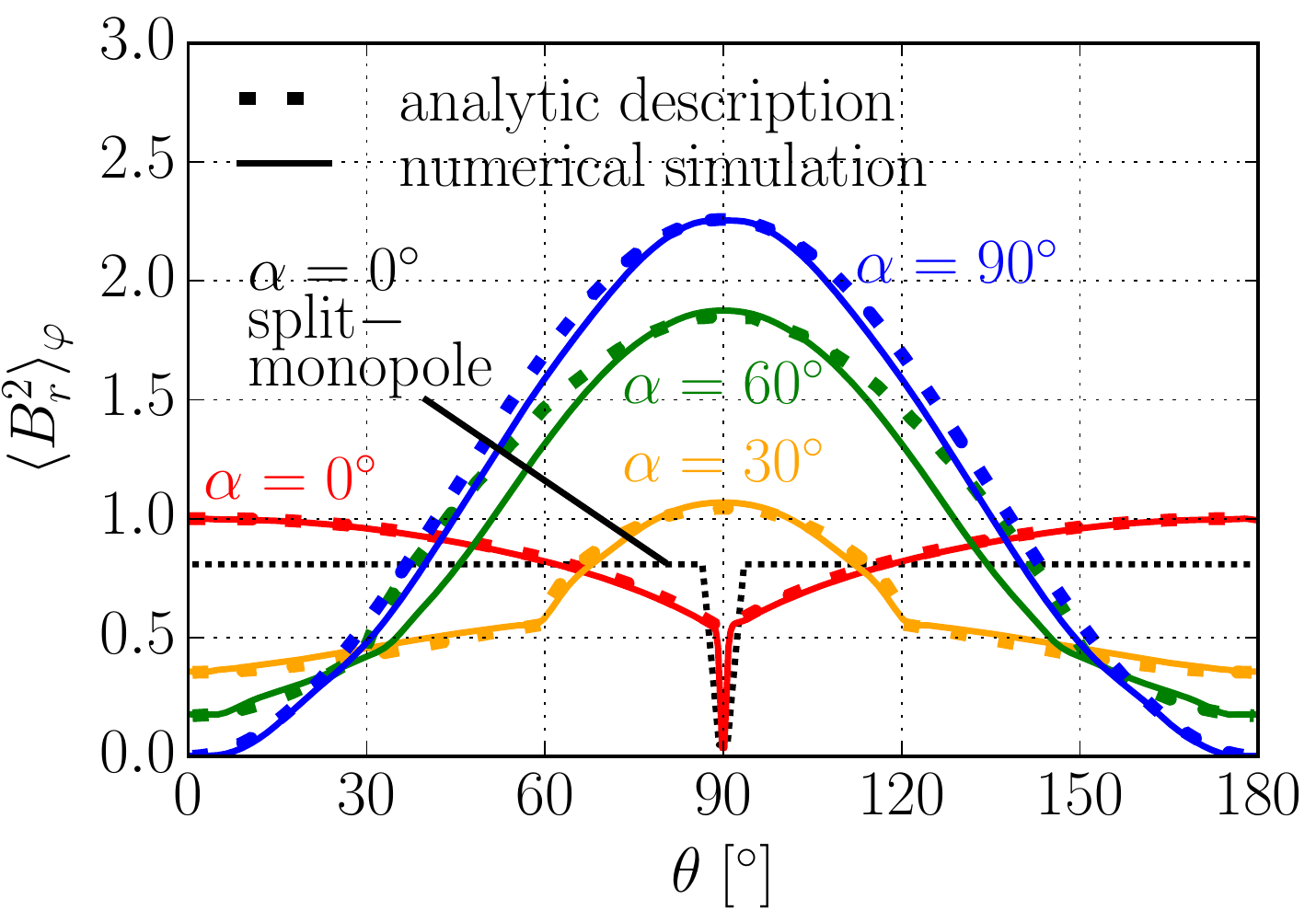}
  \end{center}
  \caption{Angular distribution of the $\langle B_r^2\rangle_\varphi$
    of the numerical force-free solution evaluated at $r = 6\Rlc$
    (solid curves) and our semi-analytical description (thick dotted curves),
    for different obliquities: $\alpha = 0, 30, 60$, and $90^\circ$,
    shown with red, orange, green, and blue lines, respectively. Black
  thin dotted line shows the aligned split-monopole solution. The
  split-monopole solution is at the same magnetic flux as the
  $\alpha=0^\circ$ aligned force-free solution, shown with the red
  lines. Clearly, solutions at all inclination angles show a
  substantial degree of non-uniformity in $B_r^2$ that is not present
  in the split-monopole solution.}
\label{fig:Brsqphianalytic}
\end{figure}

Fig.~\ref{fig:Brsqphianalytic} shows that our analytic approximation
accurately describes the azimuthally-averaged radial magnetic field
distribution, which sets the spindown power according to
eq.~\eqref{eq:P}. 
Importantly, it describes the magnetospheric structure not only for
aligned and orthogonal rotators (by construction), but also for
intermediate inclination angles. For instance, for
$\alpha = 30^\circ$, it accurately captures both the low-latitude hump
as well as the high-latitude wings in
$\avg{B_r^2}_\varphi$. Similarly, the solution provides a rather good
description of the pulsar wind at $\alpha=60^\circ$, for which it only
slightly over-predicts (by $\sim5^\circ$) the width of the magnetic
peak.  Note that the solutions at all inclination angles are
qualitatively different from the split-monopole solution, which does
not have any non-uniformity in $\theta$. For the aligned rotator, this
difference emerges from the return current present in the
magnetosphere: the return current, which is not included as part of
the split-monopole solution, causes an additional
$\vec j\times \vec B$ force, which is directed away from the midplane,
and which is compensated by the magnetic pressure gradient force,
which is caused by a non-uniformity in $B_r$ and directed
toward the midplane (see Appendix~\ref{sec:appendix}).

\begin{figure}
\begin{center}
    \includegraphics[width=1\columnwidth]{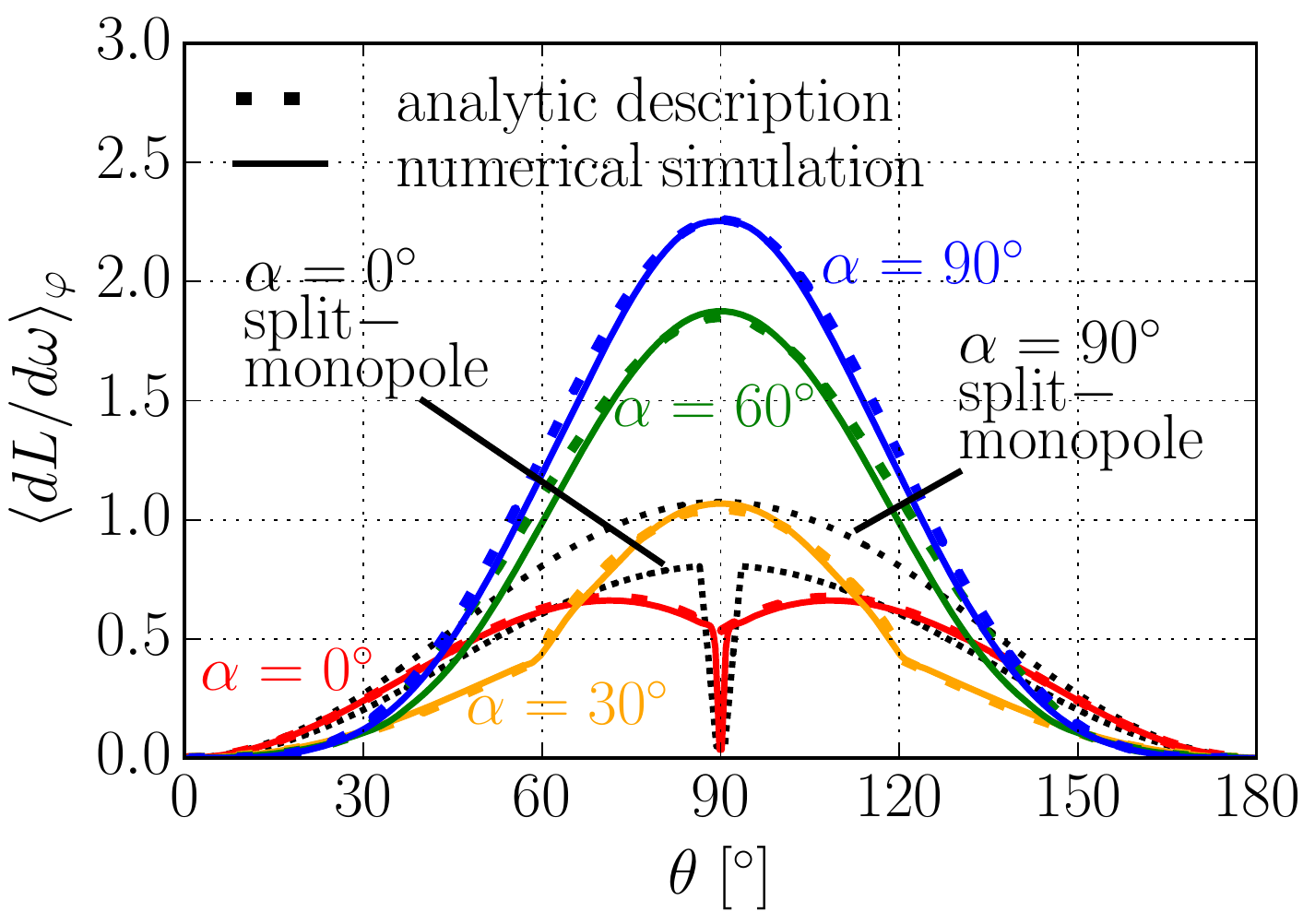}
  \end{center}
  \caption{Angular distribution of the
    $\langle dL/d\omega\rangle_\varphi$ (in arbitrary units) of the
    numerical force-free solution (solid curves) and our
    semi-analytical description (thick dotted curves) for different
    obliquities. Fine dotted black curve shows the luminosity due to an
    aligned split-monopole solution and rare dotted black curve shows
    the luminosity due to an orthogonal split-monopole solution. The
    orthogonal numerical solution has a much more pronounced peak, and
    the peak luminosity exceeds that of the split-monopole solution by a
    factor of $2$. This highlights the importance of accounting for
    the effects of non-uniformity of pulsar wind 
    on the wind luminosity and its angular distribution.}
\label{fig:dLdomegaanalytic}
\end{figure}

What kind of electromagnetic luminosity does our analytic
approximation imply? How does it compare with the simulated result?
Fig.~\ref{fig:dLdomegaanalytic} shows the distribution of luminosity
per unit solid angle,
$\avg{dL/d\omega}_\varphi\propto \avg{B_r^2}_\varphi \sin^2\theta$
(see eq.~\ref{eq:dLdtheta}), as a function of the polar angle. The
black dotted line shows the expected angular distribution for a
split-monopole model of \citet{Bogovalov},
$dL/d\omega\propto \sin^2\theta$. Whereas this dependence is broadly
representative of the aligned force-free simulation result, which we
show with the red solid line, it over-predicts the luminosity in the
equatorial region by as much as $50$\% at the expense of
under-predicting the luminosity at other angles. For other inclination
angles, the split-monopole solution under-predicts the luminosity by
as much as a factor of $2$. For instance, for $\alpha = 90^\circ$, the
split-monopole solution gives the peak value of
$\avg{dL/d\omega} \approx 1$, much smaller than the simulated peak
value is $\approx2.25$. This is despite the open magnetic flux is the
same in the split-monopole solution as in the numerical simulation.
As we discussed previously, 
this excess spindown luminosity in oblique rotators is due to the bunching of
magnetic field lines toward low-latitude regions (see
\S\ref{sec:non-unif-puls}).

\section{Discussion and Conclusions}
\label{sec:disc-concl}

In this work, we analysed the results of 3D, time-dependent numerical
simulations of oblique, plasma-filled pulsar magnetospheres in the MHD and force-free
approximations.  
Oblique plasma-filled pulsars spin down twice as fast as aligned
pulsars \citep{spit06}, i.e., the spindown luminosity $L$ varies with
the inclination angle $\alpha$ as 
\begin{equation}
L\approx L_{\rm aligned}(1+\sin^2\alpha).  \label{eq:Lvsalpha}
\end{equation}
The
reasons for such an increase are not understood.  Could this increase
be caused by the changes in the amount of open magnetic flux? Only to
a degree: we found that changes in the open magnetic flux account for
about $40$\% of the increase.  We argue that the other $60$\% of the
increase is due to a \emph{nonuniform bunching of magnetic field lines
  toward the equatorial plane}.  This redistribution allows the field
lines to extract the rotational energy more efficiently, thereby
accounting for the rest of the increase in the spindown luminosity.

To illustrate this effect, we constructed a toy model, in which we
approximated the angular distribution of the radial magnetic field,
$B_r$, by that of a dipole. Dipole magnetic field strength peaks at
the magnetic poles. As this field distribution rotates with the star,
the magnetic field lines become wound up around the axis.  The faster
the winding of a field line, the more that field line contributes to
the spindown. Since the winding is slowest at the magnetic poles
(there $v_\varphi = 0$) and fastest at the equator ($v_\varphi$ is
maximum), the field lines near the polar magnetic peak do not
contribute much to the spindown. However, in an inclined rotator, the
dipolar peak moves closer to the equatorial plane. The
field lines near the inclined peak contribute to the stellar spindown more
efficiently and lead to an increase in the spindown of an oblique
rotator relative to the aligned rotator. We showed that the spindown luminosity in
this simple model follows the dependence given in
eq.~\eqref{eq:Lvsalpha}, which illustrates the importance of magnetic
field bunching on the spindown luminosity.

The magnetic field of an actual pulsar is not dipolar.  Is there a
quantitative way of describing the non-uniformity of magnetic field in
the pulsar wind? The good news is that it is sufficient to prescribe a
single function -- a distribution of $B_r$ on a sphere of a
sufficiently large radius -- in order to fully recover the asymptotic,
3D structure of a pulsar wind.  In particular, for an aligned rotator,
the distribution of $B_r$ is axisymmetric, and we parametrised it via
a single scalar function (see eq.~\ref{eq:mhdalignfit}).

At first glance, the structure of oblique rotators appears to be
much more complicated than that of the aligned one. The good news is that
an orthogonal force-free rotator exhibits a rather featureless and
simple radial magnetic field distribution,
$B_r\propto \sin\theta\cos\varphi$, i.e., the same as of the vacuum
dipole that is put on its side (see eq.~\ref{eq:vacdipsol90})! 

Now, with these two solutions for the asymptotic pulsar wind -- for
the aligned and orthogonal plasma-filled rotators -- at hand, we combined
them into an approximate, two-component solution that describes the
simulation results at all inclination angles (\S\ref{sec:semi-analyt-solut}).

That the full solution is a combination of the two components is
particularly interesting for the following reason. The magnetic field
in the aligned component of the solution undergoes a jump in the
equatorial plane. However, the field remains smooth in the orthogonal
component.  This means that the relative
weights with which the two solutions enter the full solution tell us
about the strength of the jump in magnetic field strength and
therefore the \emph{strength of the magnetospheric current sheet}.

As the inclination angle increases, the contribution of the aligned
solution gradually decreases (eq.~\ref{eq:w1}) and the magnetospheric current sheet
weakens, i.e., the jump across the sheet in the magnetic field
decreases. The current sheet completely disappears for a force-free
orthogonal rotator, as seen in
Fig.~\ref{fig:Brsq_analyticalyaxis_map}(d),(h). This suggests that the
amount of dissipation in near-orthogonal pulsar magnetospheres might
be \emph{qualitatively} smaller than at intermediate
obliquities. This might be the reason for lower dissipation in the
current sheets found in global PIC simulations
\citep{2014arXiv1412.0673P}.
However, the absence of a jump in the magnetic field in a
force-free solution does not mean that the dissipation is entirely
absent, as found in both recent 3D MHD \citep{SashaMHD} and PIC
\citep{2014arXiv1412.0673P} simulations of oblique pulsar magnetospheres. 

Whereas force-free and MHD simulations do not include the
microphysics of the magnetospheric current sheets and reconnection, they are extremely
useful because they can be extended to extremely large distances and
give us the full, nonlinear structure of the pulsar wind
electromagnetic field that is otherwise difficult to obtain.
In particular, they
offer us a unique opportunity to determine the smooth structure of the
solution outside of the current sheets and compute the
spindown luminosity, its angular distribution, and its dependence on
the obliquity angle of the pulsar.
For instance, we find that the current sheets in our analytic
description the pulsar wind are quite unusual in that the magnitude of
$|B_r|$ undergoes a substantial jump across the
sheet.  
Since asymptotically in the pulsar wind the total (radial plus
toroidal) lab-frame magnetic pressure is tightly related to square of
the radial magnetic, $B_r^2$, via
$B^2/8\pi = (B_r^2+B_\varphi^2)/8\pi \approx
B_r^2(1+\Omega^2r^2\sin^2\theta/c^2)/8\pi$
(see eq.~\ref{eq:EB}), the lab-frame magnetic pressure also undergoes a
jump across the magnetospheric current sheet.

However, having such a jump is
at odds with the expectation that the total pressure (which is
dominated by magnetic pressure outside the sheets), must be the same on
both sides of the sheet: otherwise, the current sheet would be pushed
to the side with a lower pressure until the two pressures came to an
equilibrium. There is a flaw in this argument: since our simulations
are relativistic, we need to measure the magnetic pressure in the
\emph{fluid} frame, which is different than the pressure in the \emph{lab} frame because the pulsar
wind flows outward at a relativistic velocity, with a Lorentz factor $\gamma\gg1$ relative to the
star. Hence, it is the fluid-frame magnetic pressure,
$p_{\rm mag} \equiv b^2/8\pi \equiv B^2/8\pi\gamma^2$, 
that is continuous across the current sheet. 
How can this happen if the lab-frame
magnetic pressure is discontinuous? This can occur if the fluid
elements on the two sides of the current sheet move at
\emph{substantially different Lorentz factors}, $\gamma_1\ne\gamma_2$,
such that even though the lab-frame magnetic pressures are different,
$B_1^2/8\pi\ne B_2^2/8\pi$, the fluid-frame magnetic pressures are the same,
$B_1^2/8\pi\gamma_1^2\equiv b_1^2/8\pi \approx b_2^2/8\pi\equiv B_2^2/8\pi\gamma_2^2$.

Thus, that the square of the radial magnetic field,
$B_r^2$, undergoes a substantial jump, 
which we find both in our force-free and MHD simulations of
oblique pulsar magnetospheres as well as in our analytic model,
necessarily implies that the fluid elements on the two sides of the
current sheet \emph{move at a relativistic velocity relative to each
  other}. This relative motion is independent of the internal physics
of the current sheet and is a generic feature of the large-scale
geometry, or topology, of the pulsar magnetosphere.  This might be
interpreted as follows:
the reconnection in the current sheet is a type of \emph{driven},
rather than spontaneous, reconnection. Thus, in order to accurately
predict the properties of current sheet reconnection rate, particle
acceleration, and emission, local simulations of reconnection need to
be constructed for the specific geometry and type of discontinuity
across the current sheet in order to properly reflect the driving
exerted on the current sheet by the global magnetospheric
structure \citep{2014ApJ...783L..21S,2014PhRvL.113o5005G}. In particular, local simulations of \emph{driven}
reconnection are needed to properly represent the magnetospheric
current sheet physics.

Our analytic description of the pulsar magnetosphere provides a 3D
electromagnetic structure of a pulsar wind near and beyond the light
cylinder for both aligned and oblique pulsars.  This is useful for
constructing the models of emission from the magnetospheric current
sheet and modelling pulsar gamma-ray light curves
\citep{2013A&A...550A.101A,2014ApJ...780....3U}.  The nonuniformity of
the pulsar wind and the deviations of lateral structure from the
split-monopole expectation can be included into and will improve the
accuracy of the models of pulsar wind nebulae
\citep{2009MNRAS.400.1241C,2014MNRAS.443..547P,2014MNRAS.438..278P},
supernovae remnants, and magnetar-powered supernovae
\citep{2010ApJ...719L.204W,2010ApJ...717..245K} and core-collapse
gamma-ray bursts \citep{2011MNRAS.413.2031M}.  For all of these
systems, if an oblique rotator is a generic situation, the emerging
outflow shows smooth, sinusoidal-like variations (e.g., see
Fig.~\ref{fig:Brsq_analytical_map}) separated by jumps at the current
sheets. Because of these variations, the energy flux is more strongly
concentrated toward the edges of the emerging outflow than in the
standard split-monopole approximation for the pulsar wind.  

In the magnetar model of core-collapse gamma-ray bursts, a pulsar wind
focuses into a pair of tightly collimated jets. If jet emission
correlates with its power \citep{2012Sci...338.1445N}, variability of
the pulsar wind could translate into periodic gamma-ray variability
that is directly observable.  In the particularly clean case of an
orthogonal rotator, the emerging pulsar wind shows smooth
sinusoidal--like variation at the rotational period of the neutron
star that could lead to sinusoidal-like modulation of the high-energy
emission.  That this signature is not routinely seen in the light
curves of long-duration gamma-ray bursts suggests several
possibilities.  The central compact object could be a black hole or a
(nearly) aligned rotator and produce no periodic
variability. If the central object is an oblique
rotator, the absence of such variability might be suggestive of an
emission mechanism that washes out small time-scale variability of the
central engine.  For
instance, in sub-photospheric gamma-ray emission models (e.g.,
\citealt{2006A&A...457..763G,2008A&A...480..305G,2006ApJ...642..995P,2007A&A...469....1G,2010MNRAS.407.1033B,2013ApJ...764..157B}),
the emission is formed over a wide range of radii, which means
that the emission is sensitive only to large-scale jet structure. How
gamma-ray burst emission is produced by inhomogeneous jets warrants
further investigation.

\acknowledgements

We thank Jon Arons and Vasily Beskin for insightful discussions. 
AT was supported by NASA through Einstein Postdoctoral Fellowship
grant number PF3-140115 awarded by the Chandra X-ray Center, which is
operated by the Smithsonian Astrophysical Observatory for NASA under
contract NAS8-03060, and NASA via High-End Computing (HEC) Program
through the NASA Advanced Supercomputing (NAS) Division at Ames
Research Center that provided access to the Pleiades supercomputer, as
well as NSF through an XSEDE computational time allocation
TG-AST100040 on NICS Kraken, Nautilus, TACC Stampede, Maverick, and
Ranch. This work was supported in part by NASA grant NNX14AQ67G and Simons Foundation (grant 267233 to AS). We used Enthought Canopy Python distribution to
generate figures for this work.

\appendix{}
\section{Angular distribution of the poloidal field in the aligned pulsar magnetosphere}
\label{sec:appendix}

In this section we show that the angular structure of $B_r$ in the aligned case can be explained analytically. The force balance equation in the axisymmetric steady state can be rewritten as \citep{tch08}
\begin{equation}
\frac{B^2_p - E^2}{4\pi R_c}-({\bf R}\cdot {\bf n})\frac{B^2_{\phi}-E^2}{4\pi R} - \frac{\partial}{\partial {\bf n}}\left(\frac{B^2-E^2}{8\pi}\right) = 0,
\end{equation}
where $R_c$ is the curvature of poloidal field lines, ${\bf R}$ is cylindrical radius, and ${\bf n}$ is the unit vector along the local electric field. At large distances from the pulsar the poloidal magnetic field is mostly radial, so $\frac{\partial}{\partial {\bf n}} \approx -\frac{1}{r}\frac{\partial}{\partial \theta}$ and ${\bf R}\cdot {\bf n} \approx -\cos \theta$. With this we can rewrite the force balance as
\begin{equation}
-\frac{B^2_p - E^2}{4\pi R_c}-\cos \theta\frac{B^2_{\phi}-E^2}{4\pi R} - \frac{1}{r}\frac{\partial}{\partial\theta}\left(\frac{B^2-E^2}{8\pi}\right) = 0.
\label{eqfb}
\end{equation}
We will evaluate this equation just above the equator, thus neglecting all terms containing $\cos \theta$, and $R=r\sin\theta \approx r$. The last term in (\ref{eqfb}) can be rewritten as 
\begin{equation}
B^2- E^2 = \frac{4I^2}{c^2 r^2\sin^2\theta} + B^2_p (1 - \Omega^2 r^2 \sin \theta^2/c^2),
\end{equation}
where we used the MHD condition $E^2 = \Omega^2 R^2 B^2_p/c^2$, and the Ampere's law, which provides the relation between the toroidal field and the enclosed current $I(r,\theta)$ $B_{\phi} = 2I/cR$.
\begin{equation}
-{B^2_p}\frac{1-R^2/R^2_{LC}}{4\pi R_c} - \frac{8}{c^2 R^3} I\frac{{\rm d} I}{{\rm d} \theta} = \frac{1-R^2/R^2_{LC}}{R} \frac{\partial B^2_p}{\partial \theta}.
\end{equation}
Finally, in the wind region $R \gg R_{LC}$ we get
\begin{equation}
\frac{\partial B^2_p}{\partial \theta} = \frac{8R^2_{LC}}{c^2 R^4}I\frac{{\rm d} I}{{\rm d} \theta} - B^2_p\frac{R}{4\pi R_c}.
\label{eqder}
\end{equation}
We find that the first term dominates at $R > 2 R_{LC}$ and is negative just above the equator because of the presence of the volume-distributed return current. Finally,  we obtain $\frac{\partial B^2_p}{\partial \theta} < 0$ as it is in the simulation.

\section{Finite Magnetisation Effects}
\label{sec:finite-magn-effects}
Note that in many of the systems, such as in PWNe and SNRs, pulsar
wind propagates for many decades in radius, so that eventually the
flow becomes super-fast magnetosonic.  This happens at the fast
magnetosonic surface, at which the flow reaches Lorentz factor
$\gamma_{\rm F}\approx\gamma_{\rm max}^{1/3}$, where
$\gamma_{\rm max}$ is the terminal Lorentz factor that the outflow
would achieve if all of its electromagnetic energy were converted into
kinetic energy flux \citep{bes98,tch09}. For pulsars, theoretical
estimates suggest $\gamma_{\rm max}\sim10^4{-}10^6$
\citep[e.g.,][]{kennel_confinement_crab_1984,kennel_mhd_crab_1984}. Since in a force-free flow the Lorentz factor increases
approximately linearly with distance from the star,
$\gamma\approx r\sin\theta/\Rlc$, it crosses the fast surface at
$r_{\rm F}\sim \mu^{1/3}\Rlc \sim (20{-}100)\Rlc$. Beyond this
distance, $\gamma\sim\mu^{1/3}ln^{1/3}(r/r_{\rm F})$ (see Appendix~B
in \citealt{tch09}). The fast
surface is located at a very short distance compared to those of
interest in PWNe and SNRs, which means that finite-magnetisation, MHD
effects will set in long before the pulsar wind starts to radiate due
to the interaction with the ambient medium in these astrophysical
systems. Will this lead to large deviations away from the force-free
structure that we have considered? We are interested in the field
lines that make a small angle with the midplane,
$\theta'= \pi/2-\theta$, since it is these field lines that carry most
of the energy flux. Fortunately, once the outflow accelerates up to a large
Lorentz factor, it is extremely difficult to reorient it. In fact, the
relative change in the direction of the flow $\Delta\theta'/\theta' =
\gamma/\gamma_{\rm max}\sim \mu^{-2/3}\ln^{1/3}(r/r_{\rm F}) \ll 1$,
even for a really small $\mu\sim10^2$ and a really large
$r=10^{10}r_{\rm F}$. The only place where the finite magnetisation
MHD effects might make a difference are the polar regions of the flow,
which, however, are energetically unimportant.

\vspace{-0.3cm}
{\small
\bibliographystyle{mn2e}

}
\label{lastpage}
\end{document}